\documentclass[a4paper,11pt]{iopart}
\pdfoutput=1
\usepackage[utf8]{inputenc}
\usepackage[english]{babel}

\usepackage{graphicx}
\usepackage{bm}
\usepackage{braket}
\usepackage{amssymb}
\usepackage{cite}
\usepackage{mathrsfs}
\usepackage{xcolor}
\usepackage{hyperref}
\usepackage{enumerate}
\usepackage[toc,title]{appendix}
\usepackage{verbatim}
\usepackage[margin=30pt, bf, font=footnotesize, center, justification=justified]{caption}[2004/07/16]
\usepackage{xcolor}
\usepackage{subfigure}
\usepackage{braket}
\usepackage{xspace}
\usepackage{bbm}
\usepackage{mathdots}
\usepackage{stackrel}
\usepackage{mathtools}
\usepackage{bbm}
\usepackage[T1]{fontenc}               
\usepackage{mathrsfs}
\usepackage{appendix}
\usepackage{fancyhdr}
\usepackage{amsmath,                   
            amssymb,                   
            amsthm}                    

\usepackage{setspace} 

\def\Tr{\mathop{\mbox{\normalfont Tr}}\nolimits}

\hypersetup{colorlinks,bookmarksopen,bookmarksnumbered,
citecolor=red,
linkcolor=blue,
pdfstartview=green,
urlcolor=teal}

\catcode`@=11 
\renewcommand\tableofcontents{%
  \section*{\contentsname}%
  \@starttoc{toc}%
}
\catcode`@=12

\begin{document}

\title[Symmetry-resolved entanglement in a long-range system
]
{Symmetry-resolved entanglement in a long-range free-fermion chain}

\vspace{.5cm}

\author{Filiberto Ares$^1$, Sara Murciano$^1$, and Pasquale Calabrese$^{1,2}$}
\address{$^1$SISSA and INFN Sezione di Trieste, via Bonomea 265, 34136 Trieste, Italy.}
\address{$^{2}$International Centre for Theoretical Physics (ICTP), Strada Costiera 11, 34151 Trieste, Italy.}

\vspace{.5cm}

\begin{abstract}

We investigate the symmetry resolution of entanglement in the presence of long-range couplings. 
To this end, we study the symmetry-resolved entanglement entropy in the ground state 
of a fermionic chain that has dimerised long-range hoppings with power-like 
decaying amplitude --- a long-range generalisation of the Su-Schrieffer-Heeger model. 
This is a system that preserves the number of particles.
The entropy of each symmetry sector is calculated via the charged moments of the reduced density matrix. 
We exploit some recent results on block Toeplitz determinants generated by a discontinuous 
symbol to obtain analytically the asymptotic behaviour of the charged moments and of the symmetry-resolved entropies for a large subsystem. 
At leading order we find entanglement equipartition, but comparing with the short-range counterpart its breaking  occurs at a different 
order and it does depend on the hopping amplitudes.

\end{abstract}

\maketitle

\newpage

\tableofcontents 

\section{Introduction}
One of the most fundamental properties of entanglement is its behaviour with the size of the subsystem considered. 
For example, the entanglement entropy may follow an area law, a fact that initially 
motivated the study of this quantity due to its similarity to the black
hole entropy~\cite{Bombelli, srednicki}, and has eventually revealed the important role that entanglement plays in 
high-energy physics~\cite{RT, dong, maldacena, Raamsdonk}. Area law has been proven in the ground state of one-dimensional systems 
with mass gap and short-range couplings, in which the entanglement entropy saturates to an 
asymptotic value when the size of the subsystem is much 
larger than the correlation length~\cite{h-06}. When the mass gap is zero, the correlation length diverges, and 
the area law is corrected by a logarithmic term proportional to the central charge of the conformal 
field theory (CFT) that describes the low-energy spectrum of the model~\cite{cc-04, cc-09, hlw-94, vidal, vidal1, vidal2}. 
These properties make 
entanglement entropy a useful tool for investigating condensed matter systems~\cite{Amico, Laflorencie}. Nevertheless, the 
previous discussion changes and becomes more involved for systems with long-range couplings. 
For instance, in that case, the ground state entanglement entropy may display a logarithmic growth even if 
the mass gap is not zero, as occurs in the long-range Kitaev chain~\cite{dlegp-14, Ares0, dlegp-16, Ares1}. In general, other more exotic behaviours may arise, depending on the specific form of the couplings~\cite{Ares2}. 
The theoretical study of long-range systems has also been stimulated by the development of experimental
techniques that allow to simulate them in a laboratory~\cite{Defenu, Saffman, Monroe, Mivehar}.
In fact, R\'enyi entanglement entropies and generalisations have been experimentally measured in these ion-trap setups, especially out of 
equilibrium \cite{brydges-2018,ekh-20,vecd-20}.

In recent times, it became clear that another interesting aspect of entanglement concerns its relation 
with symmetries and, in particular, how entanglement is shared between the various symmetry sectors of a theory
\cite{lr-14,Goldstein,xavier}. The possibility of measuring in an experiment the internal symmetry structure of the entanglement \cite{fis,vecd-20,vecd1-20,ahyrst-21} was supported by new theoretical frameworks developed to address this problem \cite{Goldstein,xavier}. 
These progresses allowed to deal with the resolution of entanglement in various theoretical contexts such as CFTs 
\cite{Goldstein, xavier,goldstein1,mbc-21,cn-21,crc-20,c-21,cc-21,uv-21,nonabelian-21,bc-20,eimd-20}, free \cite{mdgc-20,hcc-21} and interacting integrable quantum field theories \cite{dhc-20,hcc-21b,chcc-21b},  
holographic settings \cite{bm-15,cnn-16,znm-20,wznm-20, znwm-22},  spin chains \cite{riccarda,SREE2dG,goldstein2,MDC-19-CTM,ccgm-20,lr-14,wv-03,bhd-18,bhd-19,bydm-20,mrc-20,tr-19,vecd-20,
vecd1-20,sela-21}, out-of-equilibrium contexts \cite{pbc-20,pbc-21,fg-21, pbc-22}, disordered systems \cite{trac-20,kusf-20,kusf-20b,kufs-20c} and for non-trivial topological phases \cite{clss-19,ms-20,as-20}. One of the main features that emerges from this literature is that conformal invariance forces the entanglement entropy to be equally distributed among the different sectors of a $U(1)$ symmetric theory \cite{xavier}.  Nonetheless, this equipartition can be spoiled by the non-universal lattice details. For example, the one-dimensional tight-binding model presents a term breaking equipartition at order $O((\log L)^{-2})$, with $L$ the size of the interval considered \cite{riccarda,bc-20,SREE2dG}. 
For scenarios out of equilibrium, a current-carrying steady state of fermions has been studied in \cite{fg-21}, finding that the first (subleading) term breaking equipartition may 
grow with the size of the subsystem. 
Also the quasiparticle dynamics of the symmetry-resolved entanglement after a quench has pointed out a violation of the equipartition at $O(1/L)$~\cite{pbc-20}.

However, to our knowledge, there are no results concerning the symmetry resolution of entanglement in the presence of long-range couplings. 
Natural questions are if the symmetry-resolved entanglement entropy, as the total one, presents different behaviours with the size of the subsystem, whether it satisfies
equipartition, and if yes what are the subleading terms breaking it. 
The goal of this work is to address these questions and to study how the total ground state entanglement splits into the contributions coming from the symmetry sectors 
of a gapped model with long-range couplings.

More specifically, the physical system that we will examine is the Su-Schrieffer-Heeger (SSH) model
with long-range hoppings. The short-range version, a dimerised tight-binding fermionic chain, 
was initially introduced to analyse solitons in conducting polymers such as polyacetylene~\cite{polym1,polym2}. In recent 
years, it has attracted a lot of interest because it supports two topologically distinct phases \cite{ssh1,ssh2, Asboth}: 
a topologically non-trivial phase, which presents zero energy states exponentially localised at the 
edges of the open chain, and a trivial phase, which is just an insulator without boundary modes. The 
relation between the entanglement properties and these two topological phases has been also investigated 
(see e.g. \cite{ssh3,ssh4,ssh5, Micallo}). Here we extend the SSH chain by including long-range dimerised 
hoppings whose amplitude decays with a power law --- this is a gapped $U(1)$ symmetric theory that preserves the 
number of particles. Other long-range generalisations of the SSH chain have been considered in the arena of 
topological insulators~\cite{Zhang-17, PerezGonzalez1, PerezGonzalez2, Ahmadi, Hsu}.

The manuscript is organised as follows. In Section \ref{sec:definitions}, we briefly review the basic concepts 
about symmetry-resolved entanglement entropy that we use throughout the work; we describe how to compute it from
the Fourier modes of the charged moments of the reduced density matrix, and we remind its most relevant 
properties in the tight-binding model. In Section \ref{sec:model}, we introduce the SSH chain with 
long-range hoppings and we calculate the ground state two-point correlation function. In the first
part of Section~\ref{sec:SREE}, we use the latter to obtain the charged moments in the thermodynamic limit. 
In particular, we derive their asymptotic behaviour for a large interval by applying some recent 
results on block Toeplitz determinants with discontinuous
symbols. In the second part of Section~\ref{sec:SREE}, we employ these results to analyse the symmetry-resolved entanglement entropy 
and finally find an asymptotic expression for it. We also benchmark the analytic results against exact numerical computations.  We conclude in Section \ref{sec:concl} with some remarks and discussions. In Appendix~\ref{app_large_L} we give a detailed derivation of the expansion of the Fourier modes of the charged moments.

\section{Entanglement resolution: tools and the short-range case}\label{sec:definitions}
In this section, we briefly review all the quantities of interest and the known results about the symmetry resolution in the one-dimensional tight-binding model, following Refs.~\cite{riccarda,SREE2dG}. 
Reviewing the short-range case allows us to illustrate some features of the entanglement resolution to which we can compare the results in the presence of long-range couplings.

\subsection{Basic definitions}
Let us consider a quantum system that can be divided into 
two parts, $A$ and $B$, such that its Hilbert space $\mathcal{H}$ 
factorises into those of $A$ and $B$, i.e. $\mathcal{H}=\mathcal{H}_A\otimes\mathcal{H}_B$.
If the full system is in a pure state $\ket{\Psi}\in \mathcal{H}$, then 
the entanglement between $A$ and $B$ may be quantified through the R\'enyi 
entanglement entropies,
\begin{equation}
 S_A^{(n)}=\frac{1}{1-n}\log\Tr\rho_A^n,
\end{equation}
where $\rho_A$ is the reduced density matrix, $\rho_A=\Tr_{\mathcal{H}_B}\ket{\Psi}\bra{\Psi}$.
In particular, the limit $n\to 1$ of $S_A^{(n)}$ gives the (von Neumann) entanglement entropy.

Here we will further assume that the system has an internal $U(1)$ symmetry.
We will denote by $Q$ the charge operator that generates such symmetry. For 
a bipartite system, the total charge is the sum of the charge in $A$ and $B$ 
and, therefore, $Q=Q_A+Q_B$. If the state of the full system $\ket{\Psi}$ is an eigenstate of $Q$, 
i.e. $Q\ket{\Psi}=q\ket{\Psi}$, one finds that $[\rho_A, Q_A]=0$. This fact implies
that the reduced density matrix $\rho_A$ admits the following block diagonal
decomposition into charge sectors,
\begin{equation}\label{rho_A_charge_dec}
 \rho_A=\bigoplus_{q}p_{A, q} \rho_{A, q}.
\end{equation}
Here $q$ runs over the eigenvalues of $Q_A$ and $p_{A,q}$ is the probability of finding the
subsystem $A$ in the sector with charge $q$, i.e. $p_{A, q}=\Tr(\Pi_q \rho_A)$, with $\Pi_q$
the projector onto the eigenspace of $Q_A$ with eigenvalue $q$.

From the density matrices $\rho_{A, q}$ in the decomposition of Eq.~\eqref{rho_A_charge_dec},
we can define the symmetry-resolved R\'enyi entropies
\begin{equation}
 S_{A, q}^{(n)}=\frac{1}{1-n}\log\Tr(\rho_{A, q}^n),
\end{equation}
which measure the entanglement between $A$ and $B$ when the subsystem $A$ is in 
the sector with charge $q$.

Here we will compute $S_{A, q}^{(n)}$ by following the approach of Refs.~\cite{Goldstein, xavier}. To this end,
we have to introduce the charged moments of $\rho_A$,
\begin{equation}\label{def_charged_mom}
 Z_A^{(n)}(\alpha)\equiv \Tr\left(\rho_A^n e^{i\alpha Q_A}\right).
\end{equation}
In Refs.~\cite{Goldstein, xavier}, the authors find that symmetry-resolved entropies $S_{A, q}^{(n)}$ can 
be obtained from the Fourier modes of the charged moments of $\rho_A$,
\begin{equation}\label{fourier_charged_mom}
 \mathcal{Z}_{A, q}^{(n)}\equiv \Tr(\Pi_q\rho_A^n)=\frac{1}{2\pi}\int_{-\pi}^{\pi}Z_A^{(n)}(\alpha)e^{-iq\alpha}d\alpha,
\end{equation}
such that
\begin{equation}\label{symm_res_ent_fourier}
 S_{A, q}^ {(n)}=\frac{1}{1-n}\log\left[\frac{\mathcal{Z}_{A, q}^{(n)}}{\left(\mathcal{Z}_{A, q}^{(1)}\right)^n}\right].
\end{equation}

\subsection{Review of the symmetry-resolved entanglement in the tight-binding model}\label{sec:U1}
The formalism explained above has been used to study the symmetry resolution of entanglement in the ground state 
of the tight-binding model
\begin{equation}\label{tbm_ham}
 H=-\sum_{n=1}^N c_n^\dagger c_{n+1}+h.c.,
\end{equation}
where $c_n^\dagger$, $c_n$ are the usual fermionic creation and annihilation  operators
that satisfy the canonical anticommutation relations,
\begin{equation}
 \{c_n^\dagger, c_m\}=\delta_{nm},\quad  \{c_n, c_m\}=\{c_n^\dagger, c_m^\dagger\}=0.
\end{equation}
This is a critical, i.e. zero mass gap, chain that preserves the particle number, $Q=\sum_{n=1}^N c^{\dagger}_n c_n$. Here we 
review the main results for $A$ being an interval of $L$ contiguous sites found in Refs.~\cite{riccarda,SREE2dG}
by exploiting the generalized Fisher-Hartwig conjecture~\cite{Basor, Basor2, JinKorepin, CalabreseEssler}. 
In order to compare with the long-range case, it will be enough to present the leading terms in $L$, 
although this conjecture allows to compute the subleading terms too.  
The asymptotic behaviour of the charged moments, defined in Eq.~\eqref{def_charged_mom}, reads
\begin{equation}\label{eq:chargedmom}
\log	Z^{(n)}_A(\alpha) =  i\frac{\alpha L}{2}-\left[\frac{1}{6}\left(n-\frac{1}{n} \right)+\frac{\alpha^2}{2\pi^2 n} \right] \log L+O(1),  
\end{equation}
where $\alpha \in [-\pi,\pi]$. Let us remark that the coefficient of the logarithmic term is quadratic in $\alpha$. 
This coefficient is universal in the sense that it does not depend on the couplings of the Hamiltonian 
(e.g. it is not affected  if we add a chemical potential in Eq.~\eqref{tbm_ham}) and it can be derived from the underlying CFT
by a generalisation of the replica trick method~\cite{Goldstein}.
Given Eq.~\eqref{eq:chargedmom}, the Fourier transformation of Eq.~\eqref{fourier_charged_mom} can 
be evaluated by the saddle point approximation at large $L$ to get
\begin{equation}
	\label{eq:ressaddleU1}
\mathcal{Z}_{A, q}^{(n)} \, =
Z_A^{(n)}(0) \frac{(\pi n)^{1/2} }{(2\log L)^{1/2}}  e^{-\frac{n(q-L/2)^2\pi^2}{2\log(L)}}+O(1/\log^{3/2} L).
\end{equation}
This result is a Gaussian function of the charge $q$ with mean $L/2$, i.e. the average number of particles 
in subsystem $A$, and variance $\log L/(\pi^2 n)$.  
Plugging this expression into Eq. (\ref{symm_res_ent_fourier}) leads to the 
symmetry-resolved R\'enyi entanglement entropy,
\begin{equation}\label{eq:intro}
 S_{A, q}^ {(n)}=
 S_{A}^ {(n)}-\frac{1}{2}\log\log L +\frac{1}{2} \frac{\log n}{1-n}+\frac{1}{2}\log\frac{\pi}{2}+o(1) .
\end{equation}
Since the Gaussian factor in Eq.~\eqref{eq:ressaddleU1} has a variance proportional
to $1/n$, it cancels when is inserted in the ratio
$\mathcal{Z}_{A, q}^{(n)}/(\mathcal{Z}_{A, q}^{(1)})^n$. This is the only term in Eq.~\eqref{eq:ressaddleU1} 
that depends on the charge $q$ and, therefore, one obtains that the symmetry-resolved entropy 
is equally distributed among the different charge sectors at leading order in $L$. 
In order to find the first term breaking the equipartition, the knowledge of the $O(1)$ 
terms in Eq. \eqref{eq:chargedmom} and their dependence on $\alpha$ are necessary, as 
it has been showed in Ref.~\cite{riccarda}. The effect of these terms is to renormalise 
the variance of the Gaussian in Eq. \eqref{eq:ressaddleU1} by a constant, 
that yields a correction in Eq. \eqref{eq:intro} proportional to $(q-L/2)^2/(\log L)^2$, 
which breaks the equipartition.

\section{Su-Schrieffer-Heeger model with long-range hoppings}\label{sec:model}
We would like to analyse the symmetry-resolved entanglement entropy in 
a system with long-range couplings. We will in particular consider the 
following fermionic chain with dimerised long-range hoppings
\begin{equation}\label{lr_ssh}
 H=-\sum_{n=1}^N\sum_{l=1}^{N/2} J_l \frac{1+(-1)^n\delta}{2} c_n^\dagger c_{n+l}+h.c.
\end{equation}
and periodic boundary conditions $c_{n+N}=c_n$.
This Hamiltonian commutes with the particle number operator $Q$. 
Notice that the hoppings extend to the whole chain, connecting sites separated
by any arbitrary distance. Therefore, in the thermodynamic limit $N\to\infty$, 
the range of the couplings is infinity.
We are going to consider the case where the hopping amplitude decays with 
the distance between sites as a power law, i.e.
\begin{equation}
 J_l=\begin{cases}
             (l+1)^{-\nu}, \quad l \,\, \mbox{odd}\\
             0, \quad l \,\, \mbox{even}.
            \end{cases}.
\end{equation}
The parameter $\nu\ge 0$ characterises the dumping of the
hopping with the distance. For simplicity, we only allow 
hoppings between even and odd sites. In order to enrich the later discussion
on the symmetry-resolved entanglement, we have taken dimerised 
hoppings and, therefore, the hopping amplitudes between even-odd and 
odd-even sites differ when $\delta\neq 0$. Thus the Hamiltonian is 
invariant under two-site translations. If $\delta=\pm 1$, the chain is
fully dimerised and there is not hopping between even-odd (odd-even) 
sites. In the case $\delta=0$, the chain is homogeneous and 
corresponds to the long-range analogue of the tight-binding model in Eq.~\eqref{tbm_ham}.

\begin{figure}[t]
\centering
 \begin{minipage}{0.5\linewidth}
 \centering 
 \includegraphics[width=\textwidth]{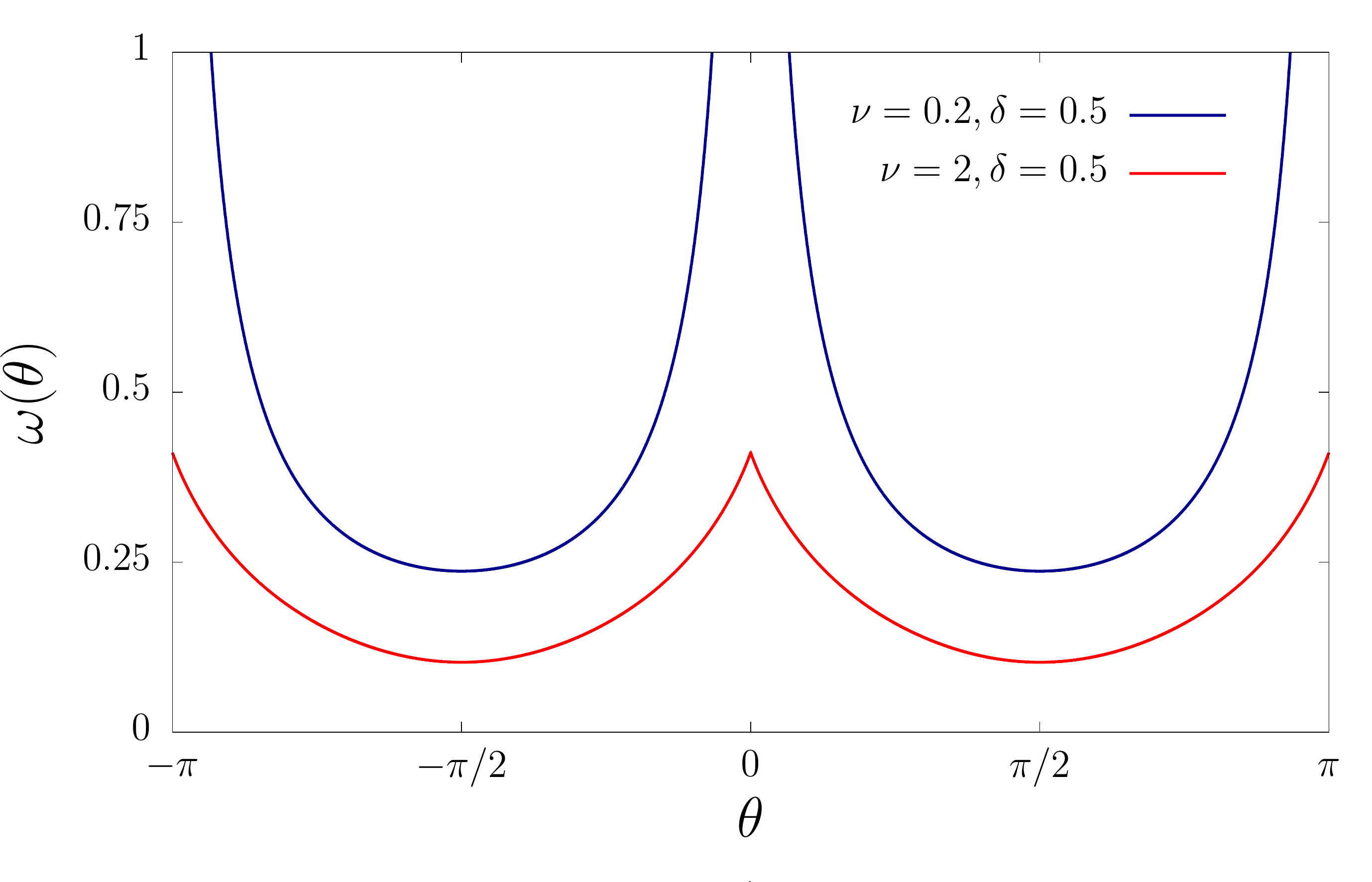}
\end{minipage}
\caption{Dispersion relation in Eq.~\eqref{disp_rel} of the long-range SSH model
for two different values of the dumping exponent $\nu$ and fixed dimerisation parameter 
$\delta=0.5$. As we explained in the main text, $\omega(\theta)$ diverges at $\theta=0,\pm \pi$ 
when $0\leq \nu<1$, while for $\nu>1$ it is finite for any $\theta$.}  \label{fig:disp_rel}
\end{figure}

This Hamiltonian can be easily diagonalised by performing 
a Fourier plus a Bogoliubov transformation. In the diagonal 
basis, it reads
\begin{equation}\label{diag_Ham}
 H=\sum_{k=0}^{N-1} \omega(\theta_k) \left(d_k^\dagger d_k-\frac{1}{2}\right),
\end{equation}
where $\theta_k=2\pi k/N$ and  $\omega(\theta)$ is the dispersion relation. 
In the thermodynamic limit $N\to\infty$, we replace $\theta_k$ by a 
continuous variable $\theta\in[-\pi, \pi]$ and the dispersion relation 
can be written in the form
\begin{equation}\label{disp_rel}
 \omega(\theta)=\sqrt{F(\theta)^2-G(\theta)^2},
\end{equation}
with 
\begin{equation}\label{Bog_F}
 F(\theta)=\frac{1}{2^{\nu+1}}\left[e^{-i\theta}{\rm Li}_\nu(e^{i 2\theta})
 +e^{i\theta}{\rm Li}_\nu(e^{-i2\theta})\right], 
\end{equation}
and
\begin{equation}\label{Bog_G}
 G(\theta)=-\frac{\delta}{2^{\nu+1}}\left[e^{-i\theta}{\rm Li}_\nu(e^{i2\theta})
 -e^{i\theta}{\rm Li}_\nu(e^{-i2\theta})\right].
\end{equation}
Here ${\rm Li}_\nu(z)$ denotes the polylogarithm function of order $\nu$~\cite{NIST}. 
The appearance of this function is due to the infinite-range hoppings 
with power-law decaying amplitude. The properties 
of the polylogarithm with the dumping exponent $\nu$ originate the features of the long-range SSH model that differ 
from the short-range systems.
In particular, ${\rm Li}_\nu(z)$ diverges
at $z=1$ when $0\leq \nu <1$, while it has a finite value for $\nu>1$. This will be
fundamental in the later analysis of the symmetry-resolved entanglement. In the 
dispersion relation, such behaviour of the polylogarithm causes it to diverge at 
$\theta=0,\pm \pi$ when $0\leq \nu<1$, while it is always finite for $\nu>1$, as we 
illustrate in Fig.~\ref{fig:disp_rel}.
For $\delta>0$ and $\nu\geq 0$, the dispersion relation is always strictly positive, 
$\omega(\theta)>0$, and therefore the mass gap is not zero. The gap closes when $\delta=0$,
the non-dimerised chain, as $\omega(\theta)$ vanishes at $\theta=\pm \pi/2$. In any case, since the dispersion 
relation is non-negative, the ground state of $H$ is the Bogoliubov vacuum, 
$\ket{0}$, defined by the property $d_k\ket{0}=0$, for all $k$. 

We would like to remark that we have written the diagonal Hamiltonian of Eq.~\eqref{diag_Ham}
in terms of a single dispersion relation, while Hamiltonians invariant under
two-site translations are usually expressed in terms of two dispersion relations. 
This depends on the choice of the domain for the momentum. In the thermodynamic limit, 
if we assume that $\theta$ takes values in the interval $\theta\in[-\pi, \pi]$,
as it is our case, then the diagonal Hamiltonian can be expressed in terms of a single positive dispersion
relation $\omega(\theta)$. On the other hand, if we restrict $\theta$ to the reduced Brillouin zone $\theta\in[-\pi/2, \pi/2]$,
then the Hamiltonian splits into two dispersion relations with opposite sign, $\pm \omega(\theta)$. These two equivalent
representations are related by mapping the modes with momentum $\theta\in[-\pi/2, \pi/2]$ and negative energy $-\omega(\theta)$
into the modes with momentum $\theta+\pi$ and positive energy $\omega(\theta)$ via a particle-hole transformation.

As the Hamiltonian is quadratic, the ground state two-point 
correlation functions are the only ingredient that we will need to 
calculate exactly the (symmetry-resolved) entanglement entropy in such 
state~\cite{Peschel}. Since the particle number is conserved, we have 
$\bra{0}c_n^\dagger c_m^\dagger \ket{0}=\bra{0} c_n c_m \ket{0}=0$, while it 
will be useful to arrange the correlations of the form $\bra{0}c_n^\dagger c_m\ket{0}$ 
in a $N\times N$ matrix $V$ with entries
\begin{equation}\label{correl_V}
 V_{l, l'}=2\left\langle 0 \left| \left(\begin{array}{c} c_{2l}^\dagger \\ c_{2l+1}^\dagger \end{array}\right)
 \left(c_{2l'}, c_{2l'+1}\right)\right| 0 \right\rangle-\delta_{l, l'},
 \quad l, l'=1, \dots, N/2.
\end{equation}
In the thermodynamic limit $N\to\infty$, $V$ is a block Toeplitz matrix,
\begin{equation}
 V_{l,l'}=\frac{1}{2\pi}\int_0^{2\pi} \mathcal{G}(\theta)e^{i\theta(l-l')}d\theta, 
\end{equation}
generated by the $2\times 2$ symbol
\begin{equation}\label{symbol_G}
 \mathcal{G}(\theta)=
 \left(\begin{array}{cc} 0 & e^{i\left(\frac{\theta}{2}-2\xi(\theta)\right)} \\ 
                         e^{-i\left(\frac{\theta}{2}-2\xi(\theta)\right)} & 0
                         \end{array}\right),
\end{equation}
where
\begin{equation}
 \cos(2\xi(\theta))=\frac{F(\theta/2)}{\sqrt{F(\theta/2)^2-G(\theta/2)^2}},
 \quad 
 \sin(2\xi(\theta))=\frac{iG(\theta/2)}{\sqrt{F(\theta/2)^2-G(\theta/2)^2}},
\end{equation}
and the functions $F(\theta)$ and $G(\theta)$ are defined in Eqs.~\eqref{Bog_F}
and \eqref{Bog_G}. Note that the $2\times 2$ block structure of the matrix $V$ is 
a consequence of the two-site translational invariance of the chain.

\section{Symmetry-resolved entanglement entropy}\label{sec:SREE}
In this section, we focus on the calculation of the symmetry-resolved entanglement 
entropies in the ground state $\ket{0}$ of the long-range SSH model. 
We start by computing exactly the charged moments for this model and then, from their Fourier transforms, we evaluate the symmetry-resolved entropies.

\subsection{Charged moments}
Since the long-range SSH model is a system described by a quadratic fermionic 
Hamiltonian, the reduced density matrix $\rho_A=\Tr_{\mathcal{H}_B}\ket{0}\bra{0}$ 
satisfies the Wick theorem and its charged moments, defined in Eq.~\eqref{def_charged_mom}, 
can be computed from the two-point correlation matrix $V$ introduced in Eq.~\eqref{correl_V}. 
More specifically, see Refs.~\cite{Peschel, Goldstein, riccarda},
\begin{equation}\label{det_Z_alpha}
 Z_{A}^{(n)}(\alpha)=\det\left[\left(\frac{I+V_A}{2}\right)^ne^{i\alpha}
 +\left(\frac{I-V_A}{2}\right)^n\right],
\end{equation}
where $V_A$ denotes the restriction of $V$ to subsystem $A$. We will employ
this expression to compute numerically the charged moments and the 
symmetry-resolved entanglement entropies by diagonalizing the correlation 
matrix $V$.

If we take into account that the eigenvalues of $V_A$ lie on the real interval
$[-1, 1]$ and we use the residue theorem, then the previous expression can be rewritten as
the contour integral~\cite{JinKorepin, riccarda}
\begin{equation}\label{contour_Z_alpha}
 \log Z_{A}^{(n)}(\alpha)=\frac{1}{2\pi i}\lim_{\varepsilon\to 1^+}
 \oint_{\mathcal{C}}f_n(\lambda/\varepsilon, \alpha)\frac{d}{d\lambda}\log D_{A}(\lambda)d\lambda,
\end{equation}
where the integration contour $\mathcal{C}$ encloses the interval $[-1, 1]$, 
\begin{equation}
 f_n(\lambda,\alpha)=\left(\frac{1+\lambda}{2}\right)^n e^{i\alpha}
 +\left(\frac{1-\lambda}{2}\right)^n,
\end{equation}
and $D_A(\lambda)$ denotes the characteristic polynomial of $V_A$, i.e. 
$D_A(\lambda)=\det(\lambda I-V_A)$.

The previous discussion is valid for any subsystem $A$. Here we will focus 
on one consisting of a single interval of $L$ contiguous sites.
In this case, the restriction $V_A$ is a $L\times L$ block Toeplitz matrix
with symbol the $2\times 2$ matrix $\mathcal{G}(\theta)$ of Eq.~\eqref{symbol_G}. 
To deduce the large $L$ behaviour of $D_A(\lambda)$ and, therefore,
of the charged moments $Z_{A}^{(n)}(\alpha)$, we will apply the 
results on the asymptotic behaviour of block Toeplitz determinants obtained 
in Ref.~\cite{Ares1}. In particular, if the symbol $\mathcal{G}_\lambda(\theta)
=\lambda I-\mathcal{G}(\theta)$ satisfies $\det\mathcal{G}_\lambda(\theta)\neq 0$
and is a piecewise continuous function in $\theta$ with jump discontinuities 
at $\theta=\theta_1,\dots,\theta_R$, then 
\begin{equation}\label{asymp_log_D_A}
 \log D_A(\lambda)= \frac{L}{4\pi}\int_0^{2\pi}\log\det\mathcal{G}_\lambda(\theta)d\theta 
 +\frac{\log L}{4\pi^2}\sum_{r=1}^R \Tr[\log \mathcal{G}_{\lambda,r}^-(\mathcal{G}_{\lambda,r}^+)^{-1}]^2+O(1),
\end{equation}
where $\mathcal{G}_{\lambda,r}^\pm$ are the lateral limits of $\mathcal{G}_\lambda(\theta)$ in 
the jump discontinuity at $\theta=\theta_r$,
\begin{equation}
 \mathcal{G}_{\lambda,r}^\pm=\lim_{\theta\to\theta_r^\pm}\mathcal{G}_\lambda(\theta).
\end{equation}
Eq.~\eqref{asymp_log_D_A} is a generalisation of the Fisher-Hartwig conjecture for block Toeplitz determinants ~\cite{FH, Basor3}.
If the symbol $\mathcal{G}_\lambda(\theta)$ has continuous entries in $\theta$, then there is no logarithmic term in the asymptotic 
expansion of $\log D_A(\lambda)$, and Eq.~\eqref{asymp_log_D_A} simplifies to
\begin{equation}
 \log D_A(\lambda)= \frac{L}{4\pi}\int_0^{2\pi}\log\det\mathcal{G}_\lambda(\theta)d\theta+O(1).
\end{equation}
This is the Szeg\H{o}-Widom theorem~\cite{Widom}.

Let us first consider the case $\delta=0$ when the chain is not dimerised and the mass gap is zero. In this situation,
the symbol $\mathcal{G}_\lambda(\theta)$ presents a jump discontinuity 
at $\theta=\pi$ due to the zeros of the dispersion relation $\omega(\theta)$. 
The lateral limits at this point are 
\begin{equation}
 \mathcal{G}_{\lambda, \pi}^{\pm}=\lambda I \pm \sigma_y,
\end{equation}
where $\sigma_y$ is the Pauli matrix. Thus, by applying the result of Eq.~\eqref{asymp_log_D_A},
we conclude that there is a logarithmic term in $\log D_A(\lambda)$, 
\begin{equation}
 \log D_A(\lambda)= \log(\lambda^2-1)L+\frac{\log L}{2\pi^2}\left(\log\frac{\lambda-1}{\lambda+1}\right)^2+O(1).
\end{equation}
If we plug this result into the contour integral of Eq.~\eqref{contour_Z_alpha},
we obtain that the charged moments for $\delta = 0$ are of the form
\begin{equation}
 \log Z_A^{(n)}(\alpha)=i\frac{\alpha L}{2}+B_n(\nu, 0, \alpha)\log L+O(1),
\end{equation}
where 
\begin{equation}\label{B_n_delta_0}
 B_n(\nu, 0, \alpha)=\frac{1}{\pi^3 i}\lim_{\varepsilon\to 1^+} 
 \oint_\mathcal{C} \frac{f_n(\lambda/\varepsilon, \alpha)}{1-\lambda^2}
 \log\left( \frac{1+\lambda}{1-\lambda}\right)d\lambda.
\end{equation}
This integral has already appeared in Ref.~\cite{riccarda},
where it is explicitly worked out. The final result is
\begin{equation}
 B_n(\nu, 0, \alpha)=-\left[\frac{1}{6}\left(n-\frac{1}{n}\right)+\frac{\alpha^2}{2\pi^2n}\right].
\end{equation}
$ B_n(\nu, 0, \alpha)$ is equal to the coefficient of the logarithmic term of the 
charged moments for the ground state of the tight-binding model, see Eq.~\eqref{eq:chargedmom}. 
This is a consequence of the fact that for $\delta=0$ the ground state is always a Fermi sea, independently of the values of $\nu$.
Therefore, when the long-range chain of Eq.~\eqref{lr_ssh}
is not dimerised, i.e. $\delta=0$, the discussion on the symmetry-resolved entropies
follows similar lines as for the short-range chain sketched in Section~\ref{sec:definitions}.

In order to get a behaviour different from the short-range case, we need to
dimerise the chain, that is, to take $\delta \neq 0$. Now the 
source of discontinuities in the symbol $\mathcal{G}_\lambda(\theta)$ 
is the polylogarithm function ${\rm Li}_\nu(z)$ that encodes the long-range hoppings. In the interval $0\leq \nu<1$,
${\rm Li}_\nu(z)$ diverges at the point $z=1$. This divergence produces a 
jump discontinuity in $\mathcal{G}_\lambda(\theta)$ at $\theta=0$ 
for $0\leq \nu<1$ and $\delta\neq 0$ with lateral limits
\begin{equation}\label{lat_lim_G}
 \mathcal{G}_{\lambda,0}^\pm=\lambda I+\cos\xi_0 \sigma_z\pm\sin\xi_0 \sigma_y,
\end{equation}
where $\sigma_y$ and $\sigma_z$ are the Pauli matrices and 
\begin{equation}
 \cos\xi_0=\frac{\sin(\pi\nu/2)}{\sqrt{\delta^2\cos^2(\pi\nu/2)+\sin^2(\pi\nu/2)}},
 \end{equation}
 and
 \begin{equation}
 \sin\xi_0=\frac{\delta\cos(\pi\nu/2)}{\sqrt{\delta^2\cos^2(\pi\nu/2)+\sin^2(\pi\nu/2)}}.
\end{equation}
According to Eq.~\eqref{asymp_log_D_A}, this discontinuity gives rise to a logarithmic 
term in $\log D_A(\lambda)$. If we plug Eq.~\eqref{lat_lim_G} into Eq.~\eqref{asymp_log_D_A}, 
we find that 
\begin{equation}\label{asymp_log_D_A_nu_0_1}
 \log D_A(\lambda)=\log(\lambda^2-1) L+b_0(\lambda) \log L + O(1),
\end{equation}
for $0\leq \nu<1$ and $\delta\neq0$, with 
\begin{equation}
 b_0(\lambda)=\frac{2}{\pi^2}\left(\log\frac{\sqrt{\lambda^2-\cos^2\xi_0}+\sin\xi_0}{\sqrt{\lambda^2-1}}\right)^2.
\end{equation}
For $\nu>1$, the polylogarithm ${\rm Li}_\nu(z)$ converges in  
all the unit circle $z=e^{i\theta}$. This implies that the symbol 
$\mathcal{G}_\lambda(\theta)$ has continuous entries in $\theta$ 
when $\nu\geq 1$ and $\delta\neq 0$. In this case, there is no logarithmic 
term in $\log D_A(\lambda)$, and
\begin{equation}
 \log D_A(\lambda)=\log(\lambda^2-1)L+O(1).
\end{equation}

In what follows, we will focus on the range $0\leq \nu <1$ and $\delta\neq 0$.
If we insert Eq.~\eqref{asymp_log_D_A_nu_0_1} in Eq.~\eqref{contour_Z_alpha},
we obtain the following asymptotic behaviour for the charged moments of $\rho_A$
\begin{equation}\label{Z_alpha_n_0_1}
 \log Z_{A}^{(n)}(\alpha)=i\frac{\alpha L}{2} + B_n(\nu, \delta, \alpha)\log L+O(1),
\end{equation}
when $0\leq \nu<1$ and $\delta\neq 0$. The coefficient $B_n(\nu, \delta, \alpha)$ 
is given by the contour integral 
\begin{equation}\label{eq:B_n1}
 B_n(\nu, \delta, \alpha)=\frac{1}{2\pi i}\lim_{\varepsilon\to 1^+} 
 \oint_\mathcal{C} f_n(\lambda/\varepsilon, \alpha)\frac{db_0(\lambda)}{d\lambda}d\lambda,
\end{equation}
which, following similar steps as in Ref.\cite{Ares1}, can be reduced to the real integral
\begin{equation}\label{B_n}
 B_n(\nu, \delta, \alpha)=\frac{4}{\pi^2}
 \int_{\cos\xi_0}^1 g_n(\lambda, \alpha)\log\left(\frac{\sqrt{1-\lambda^2}}
 {\sqrt{\lambda^2-\cos^2\xi_0}+\sin\xi_0}\right)d\lambda,
\end{equation}
where 
\begin{equation}
 g_n(\lambda, \alpha)=n\frac{(1+\lambda)^{2n-1}+\cos\alpha\left[(1+\lambda)^{n-1}(1-\lambda)^n-(1+\lambda)^n(1-\lambda)^{n-1}\right]-(1-\lambda)^{2n-1}}{(1+\lambda)^{2n}+2[(1+\lambda)(1-\lambda)]^n\cos\alpha+(1-\lambda)^{2n}}.
\end{equation}

\begin{figure}[t]
 \begin{minipage}{0.5\linewidth}
 \centering 
 \includegraphics[width=\textwidth]{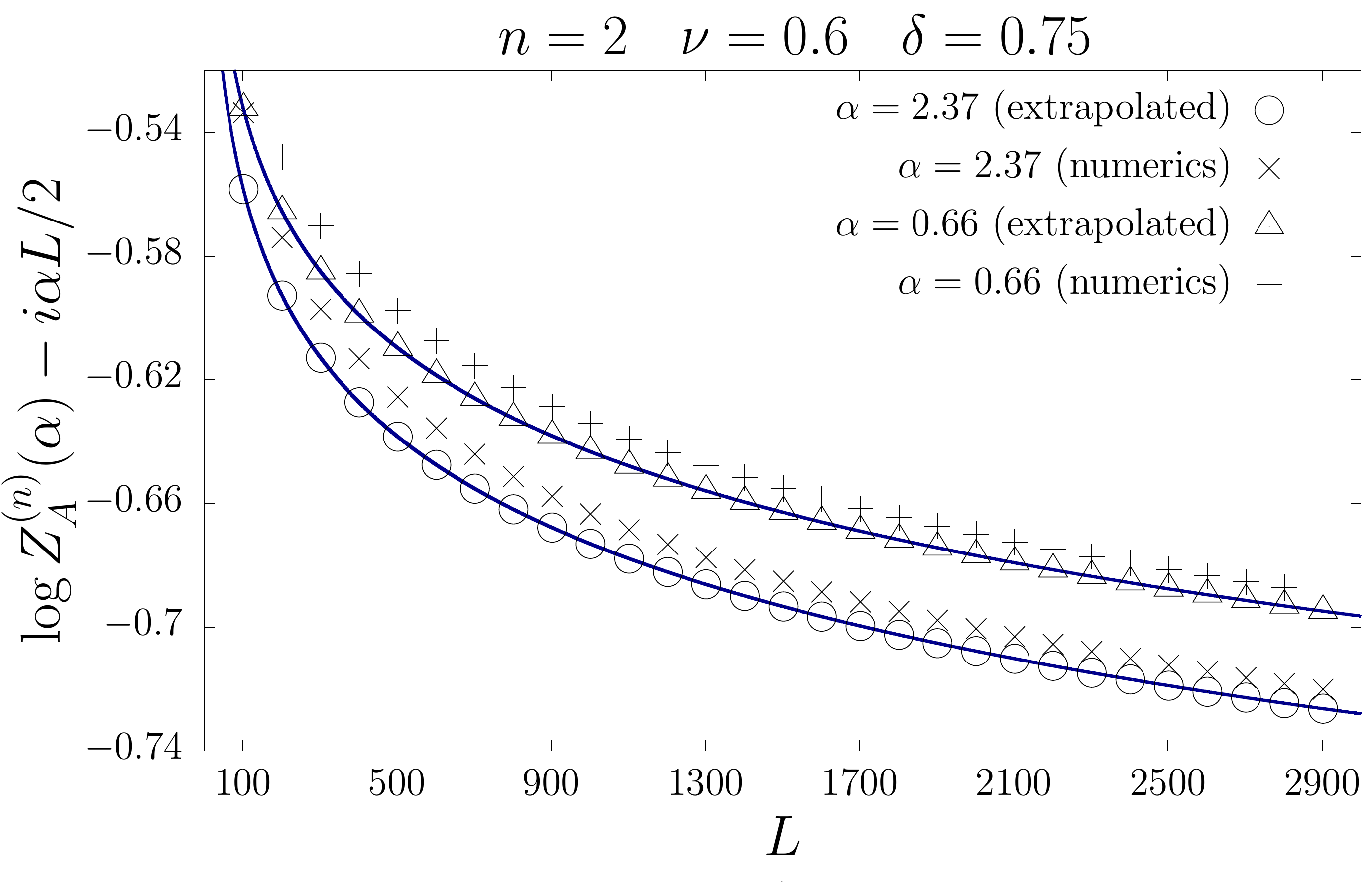}
\end{minipage}
 \begin{minipage}{0.5\linewidth}
 \centering 
 \includegraphics[width=\textwidth]{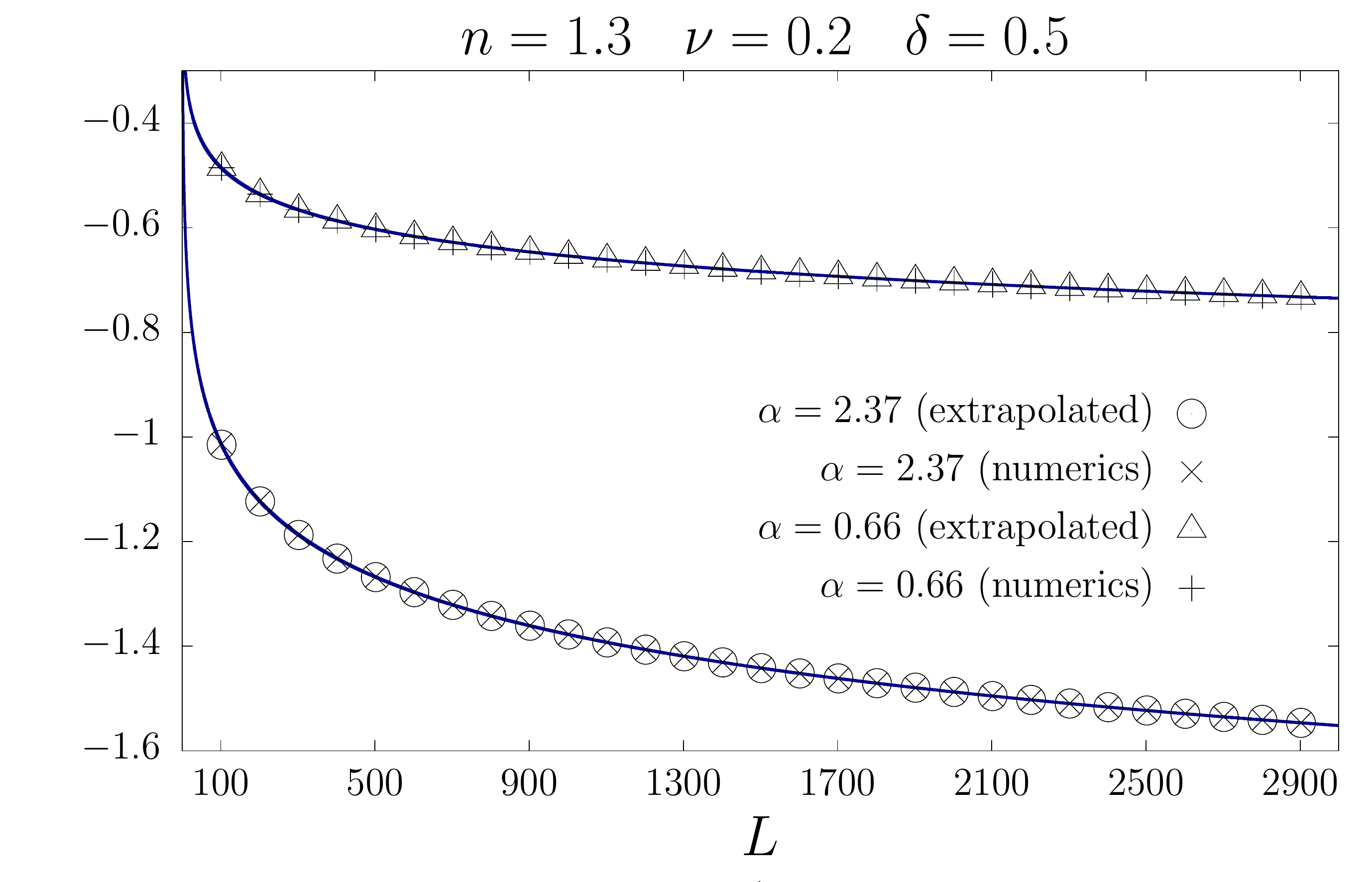}
\end{minipage}
\caption{Numerical check of the asymptotic behaviour derived in Eq.~\eqref{Z_alpha_n_0_1} for the 
charged moments of $\rho_A$ for two different sets of couplings $\nu$ and $\delta$. 
The points indicated as \textit{numerics} have been directly obtained from Eq.~\eqref{det_Z_alpha} 
by exact diagonalisation of the correlation matrix $V_A$. Using these points we have estimated the first
$O(1)$ corrections in Eq.~\eqref{Z_alpha_n_0_1}: we have subtracted from them the leading contribution 
$B_n(\nu, \delta, \alpha)\log L$ and then, with the resulting points, we have fitted the function 
$C_n(\nu, \delta, \alpha)+D_n(\nu, \delta, \alpha)L^{-D'_n(\nu, \delta, \alpha)}$. The \textit{extrapolated} points have 
been obtained from the \textit{numerics} ones by subtracting the term $D_n(\nu, \delta, \alpha)L^{-D_n'(\nu, \delta, \alpha)}$
with the values for $D_n(\nu, \delta, \alpha)$ and $D_n'(\nu, \delta, \alpha)$ found in the fit. The continuous
line represents $B_n(\nu,\delta, \alpha)\log L+C_n(\nu, \delta, \alpha)$, taking for $C_n(\nu,\delta,\alpha)$ the value
given by the fit.}\label{fig:log_charged_mom}
\end{figure}

In Fig.~\ref{fig:log_charged_mom}, we check numerically the asymptotic behaviour for the charged moments 
$Z_{A}^{(n)}(\alpha)$ predicted in Eq.~\eqref{Z_alpha_n_0_1}. We obtain an excellent agreement once the subleading corrections
are taken into account. Unfortunately, to our knowledge, there are no results in the theory
of block Toeplitz determinants that allow us to extract analytically these corrections, as a difference with the short-range case \cite{CalabreseEssler,ccp-10}. Nevertheless,
from the analysis of the numerical data, we conjecture that the first subleading terms in Eq.~\eqref{Z_alpha_n_0_1} 
are of the form $C_n(\nu, \delta, \alpha)+D_n(\nu, \delta, \alpha)L^{-D_n'(\nu, \delta, \alpha)}$. The value of the 
coefficients $C_n(\nu, \delta, \alpha)$, $D_n(\nu, \delta, \alpha)$ and $D_n'(\nu, \delta, \alpha)$
for different sets of parameters can be estimated by a fit with the numerical data, as we explain in detail 
in the caption of Fig.~\ref{fig:log_charged_mom}.

Compared with the short-range systems, the most striking feature of the charged moments in the ground state of the long-range 
SSH model is the appearance of a $\log L$ term in $\log Z_A^{(n)}(\alpha)$ when the dumping
exponent is $0\leq \nu <1$, even though the mass gap is not zero. 
On the contrary, in systems with short-range couplings, such term arises when the mass gap vanishes, as it is evident from Eq. \eqref{eq:chargedmom}. 
In the gapped short-range systems studied so far, such as for example  the complex harmonic 
chain~\cite{MDC-19-CTM} and the XXZ spin chain~\cite{ccgm-20}, there is no logarithmic term in $\log Z_A^{(n)}(\alpha)$ and 
its real part saturates to a constant in the limit $L\to\infty$, as occurs in our case when $\nu\geq1$.
The presence of this logarithmic term for gapped systems is a genuine feature of the long-range hoppings. 

\begin{figure}[t]
 \begin{minipage}{0.5\linewidth}
 \centering 
 \includegraphics[width=\textwidth]{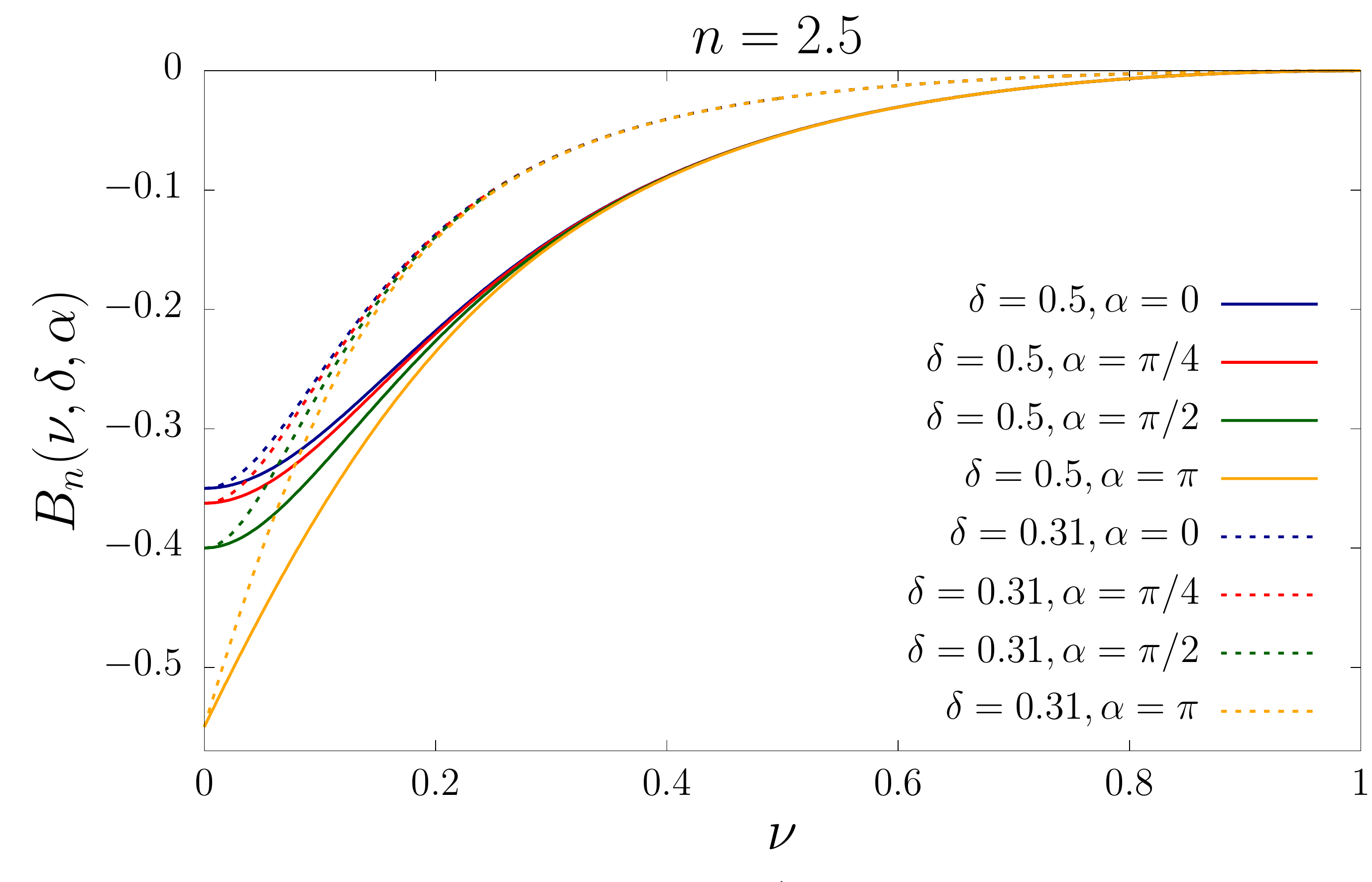}
\end{minipage}
 \begin{minipage}{0.5\linewidth}
 \centering 
 \includegraphics[width=\textwidth]{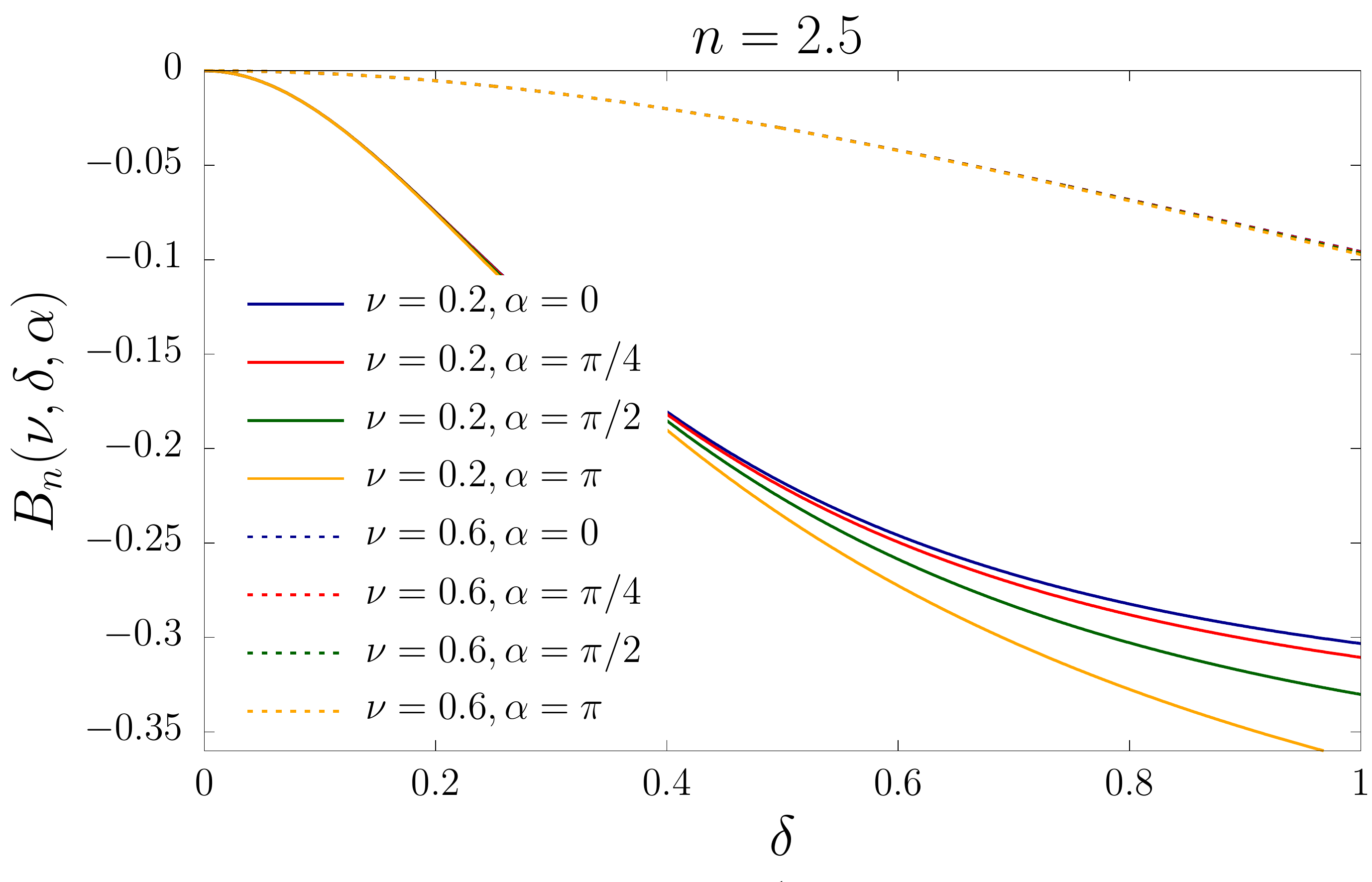}
\end{minipage}
\begin{minipage}{0.5\linewidth}
 \includegraphics[width=\textwidth]{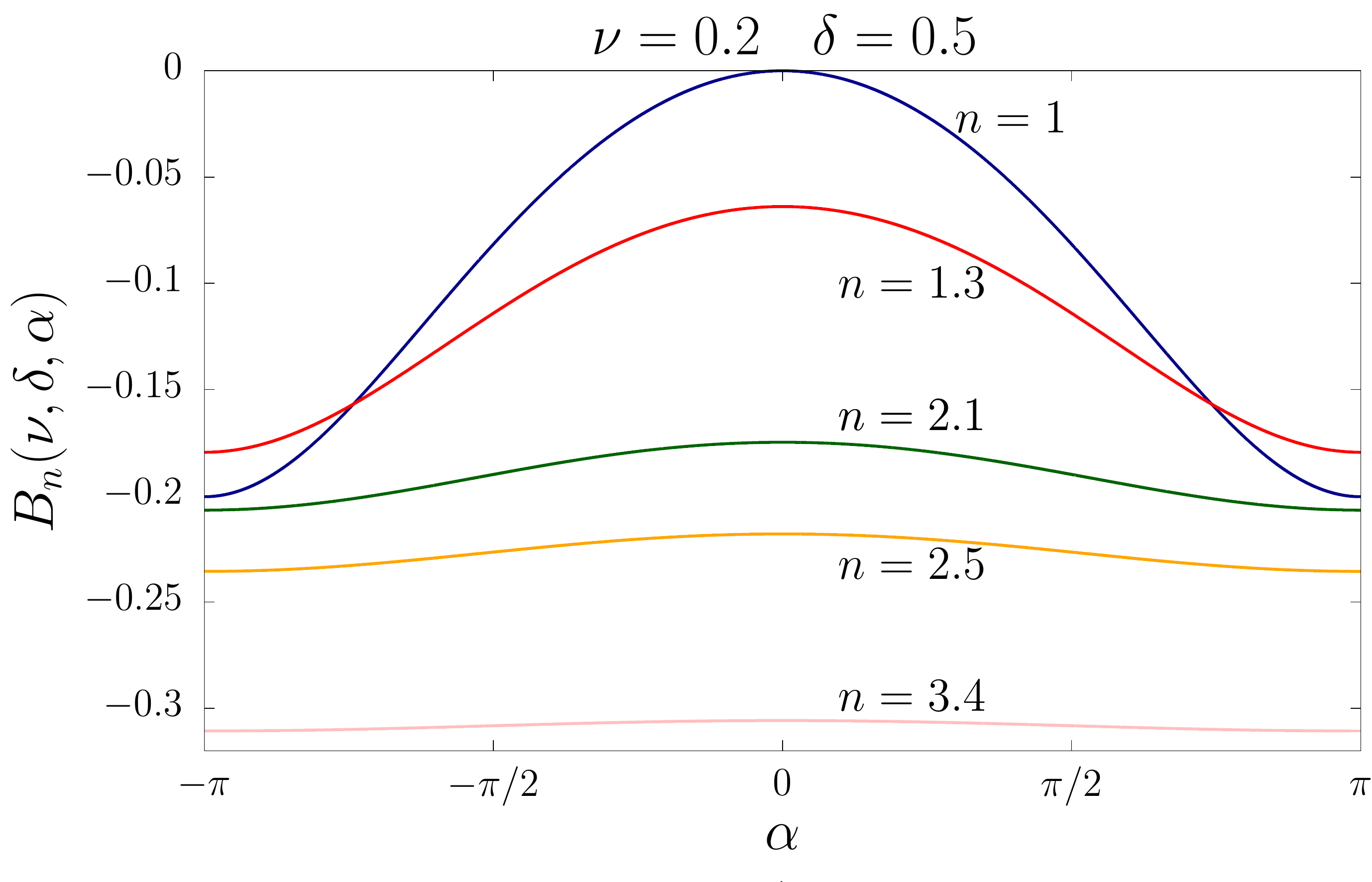}
\end{minipage}
\begin{minipage}{0.5\linewidth}
 \includegraphics[width=\textwidth]{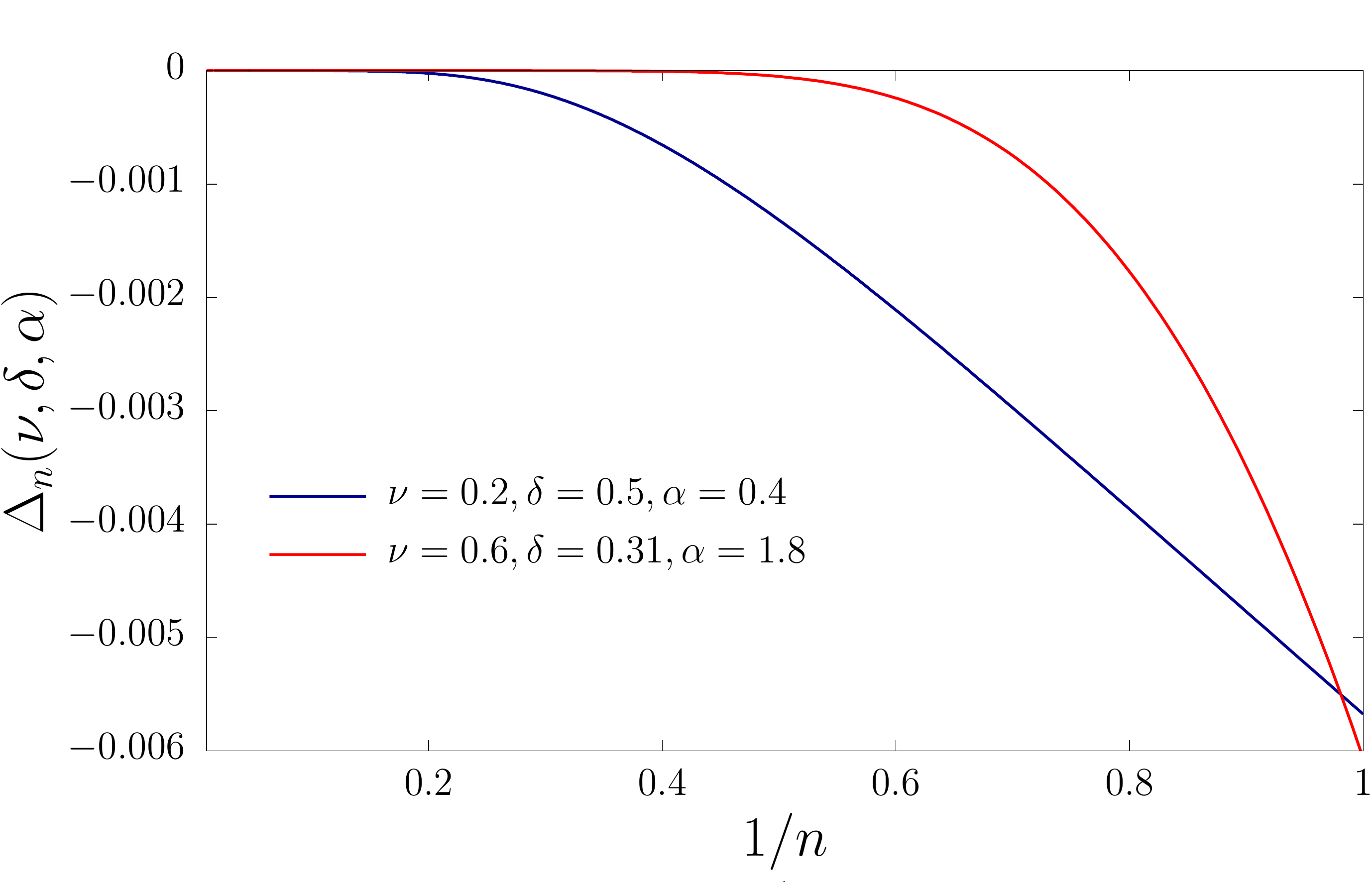}
\end{minipage}
\caption{Plot of the coefficient $B_n(\nu, \delta, \alpha)$ given in Eq.~\eqref{B_n} as 
         a function of $\nu$ (upper left panel), $\delta$ (upper right panel) and $\alpha$ (lower left
         panel). In the lower right panel, we represent $\Delta_n(\nu, \delta,\alpha)=B_n(\nu, \delta, \alpha)-B_n(\nu, \delta, 0)$
         versus the inverse of the R\'enyi parameter $n$.}\label{fig:B_n}
\end{figure}

Here the coefficient $B_n(\nu,\delta,\alpha)$ of the logarithmic term in $\log Z_A^{(n)}(\alpha)$ is 
not universal as one can see in Fig.~\ref{fig:B_n}, where we plot it 
as a function of $\nu$, $\delta$ and $\alpha$. In fact, given
the involved expression of Eq.~\eqref{B_n}, $B_n(\nu, \delta, \alpha)$ shows 
a non-trivial dependence on the different parameters. As we already remarked, it 
vanishes when $\nu\geq 1$ and, as we show in the left upper plot of Fig.~\ref{fig:B_n},   
it smoothly tends to zero in the limit $\nu\to1^-$. An interesting 
case is $\nu=0$: starting from Eq.~\eqref{lat_lim_G}, then Eq.~\eqref{eq:B_n1}
simplifies for any $\delta\neq 0$ to the same integral of Eq.~\eqref{B_n_delta_0},
obtained when $\delta=0$, and therefore
\begin{equation}\label{B_n_nu_0}
B_n(0, \delta, \alpha)=-\left[\frac{1}{6}\left(n-\frac{1}{n}\right)+\frac{\alpha^2}{2\pi^2 n}\right].
\end{equation}
Notice that this is equal to the coefficient of the logarithmic term of $\log Z_A^{(n)}(\alpha)$ in the gapless
tight-binding model reported in Eq.~\eqref{eq:chargedmom}. Nevertheless, it is important to recall that in our case the system 
is gapped if $\delta\neq0$ and, moreover, when $\nu=0$ the hopping couplings, which extend over the whole chain, do 
not decay with the distance between sites.

The coefficient $B_n(\nu, \delta, \alpha)$ presents remarkable 
properties also as a function of the parameters $\alpha$ and $n$. It will be useful to 
consider the difference
\begin{equation}\label{Delta_n}
 \Delta_n(\nu, \delta, \alpha)=B_n(\nu, \delta, \alpha)-B_n(\nu, \delta, 0).
\end{equation}
In general, $\Delta_n(\nu, \delta, \alpha)$ has the following power 
series expansion in $\alpha$ around $\alpha=0$,
\begin{equation}\label{Delta_n_power_series}
 \Delta_n(\nu, \delta, \alpha)=\sum_{j=1}^\infty \Delta_n^{(2j)}(\nu, \delta)\alpha^{2j},
\end{equation}
with $\Delta_n^{(2)}<0$. This is in contrast to the tight-binding chain reviewed in Section \ref{sec:definitions}, for which $ \Delta_n=-1/(2\pi^2 n)$  (equal to the cases $\delta=0$ and $\nu=0$ analysed in Eqs.~\eqref{B_n_delta_0} and \eqref{B_n_nu_0} respectively), i.e.
it is a quadratic function in $\alpha$, as also happens 
in the critical limit of the complex Harmonic chain~\cite{MDC-19-CTM} and in all cases present in the literature. 
Moreover, in  these short-range models, $\Delta_n$ is proportional 
to $1/n$, something that, in general, does not hold in our case, as we check in the lower right panel
of Fig.~\ref{fig:B_n}. This fact will have important consequences in the symmetry-resolved entropy of the long-range SSH.

\subsection{Symmetry resolution via Fourier transform}
Now we can derive from Eq.~\eqref{asymp_log_D_A_nu_0_1} the asymptotic behaviour 
of the Fourier tranforms of the charged moments $Z_{A}^{(n)}(\alpha)$ for $0\leq \nu<1$, 
defined in Eq.~\eqref{fourier_charged_mom}. In our case, using the results
of the previous subsection, they can be rewritten in the form
\begin{equation}\label{fourier_charged_mom_n_0_1}
 \mathcal{Z}_{A, q}^{(n)}=\frac{Z_{A}^{(n)}(0)}{2\pi}\int_{-\pi}^\pi
 e^{-i\alpha(q-L/2)}e^{\Delta_n(\nu, \delta, \alpha)\log L+O(1)}d\alpha.
\end{equation}
In this expression we have factorised the $\alpha=0$ contribution,
which behaves for large $L$ as
\begin{equation}
 Z_{A}^{(n)}(0)=e^{B_n(\nu, \delta, 0)\log L+O(1)},
\end{equation}
and eventually yields the total R\'enyi entanglement entropy,
\begin{equation}\label{full_renyi_ee_ssh}
 S_A^{(n)}=\frac{1}{1-n}\log Z_A^{(n)}(0)=
 \frac{B_n(\nu, \delta, 0)}{1-n}\log L+O(1).
\end{equation}
As we already pointed out for the charged moments, despite the mass gap of the system does not vanish, we find a logarithmic growth of the entanglement rather than the usual saturation to a constant value \cite{h-06, cc-04,ccp-10}.

\begin{figure}[t]
 \begin{minipage}{0.5\linewidth}
 \centering 
 \includegraphics[width=\textwidth]{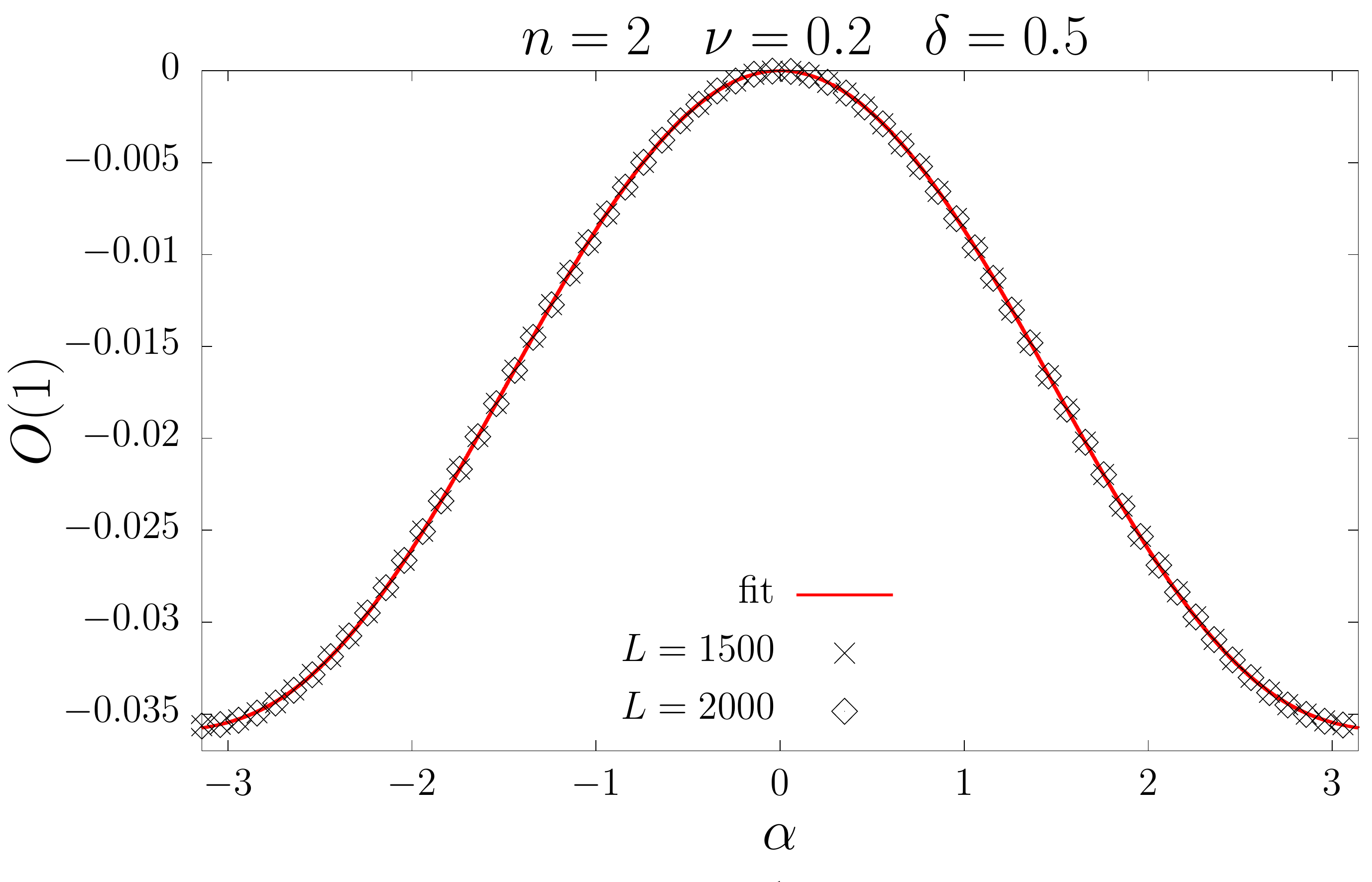}
\end{minipage}
 \begin{minipage}{0.5\linewidth}
 \centering 
 \includegraphics[width=\textwidth]{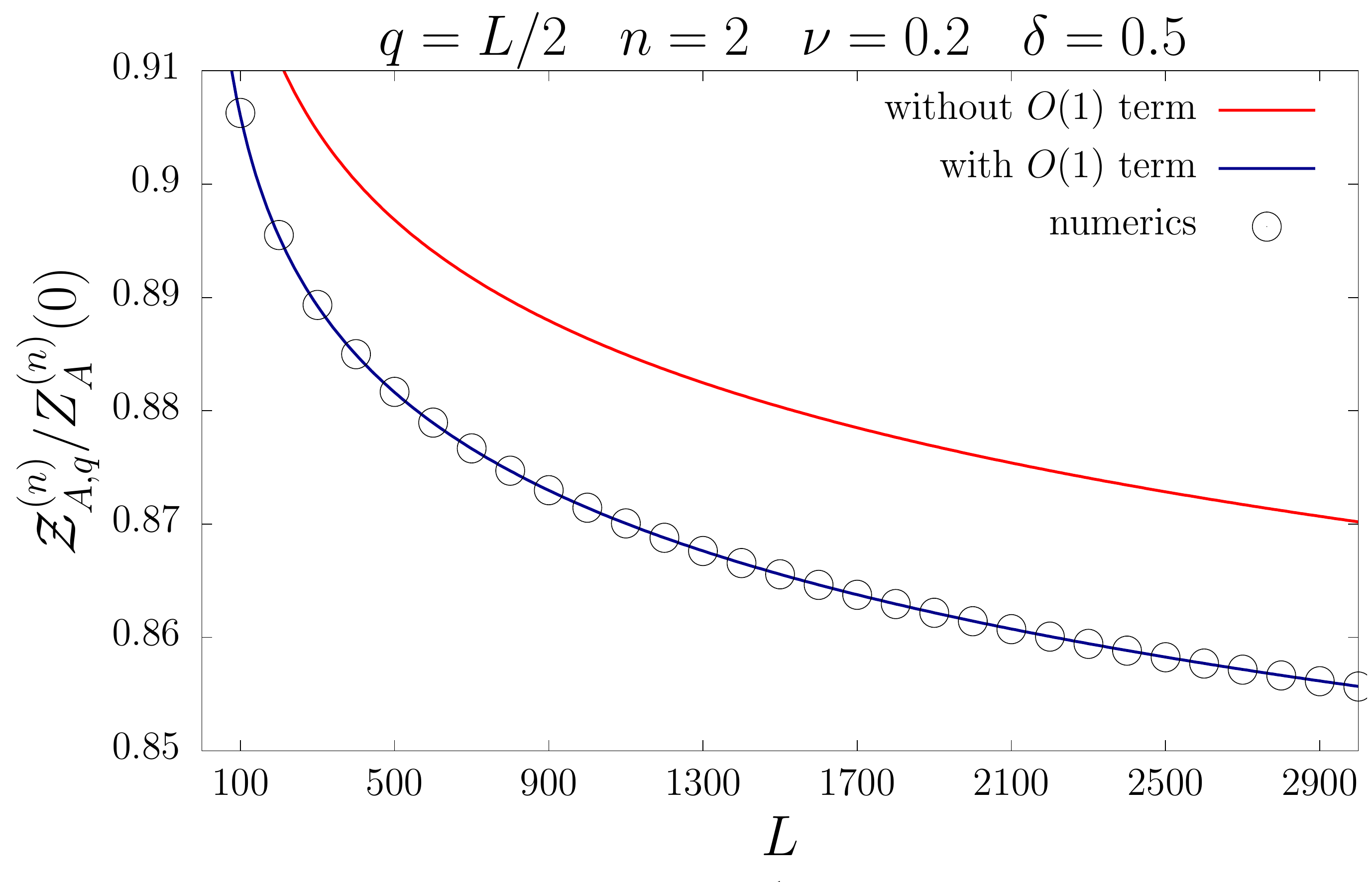}
\end{minipage}
\caption{In the left panel, we represent the $O(1)$ term of the exponent of the integrand
of Eq.~\eqref{fourier_charged_mom_n_0_1}, obtained numerically using Eq.~\eqref{det_Z_alpha} 
for two different interval lengths. We have fitted a function of the form $\sum_{j=1}^8 b_{2j} 
\alpha^{2j}$ to the points of the case $L=1500$. The continuous line represents this function using the coefficients 
arising from the fit. In the right panel, we analyse the $q=L/2$ Fourier coefficient $\mathcal{Z}_{A, q}^{(n)}$ 
of the charged moments. The points have been obtained from the diagonalisation of the correlation
matrix $V_A$ through Eq.~\eqref{det_Z_alpha}. The red line corresponds to performing the numerical
integration of Eq.~\eqref{fourier_charged_mom_n_0_1} without including the $O(1)$ terms while in the blue  
curve we have added the $O(1)$ term estimated in the fit with the $L=1500$ points of the left 
plot of the present Figure.}\label{fig:calZ}
\end{figure}

The coefficient $\Delta_n(\nu, \delta, \alpha)$ in the integrand of Eq.~\eqref{fourier_charged_mom_n_0_1} is 
the difference defined in Eq.~\eqref{Delta_n}. As we have already mentioned,
we do not have analytical methods to obtain the subleading terms in $L$ 
that appear in the exponent of this integrand. In Fig.~\ref{fig:calZ}, we have studied 
numerically the $O(1)$ term for a particular set of couplings. In general,
this term is an even function in $\alpha$ and, therefore, can be expressed
as a power series of the form $\sum_{j=1}^\infty b_{2j} \alpha^{2j}$. In the left panel 
of Fig.~\ref{fig:calZ}, we have fitted this function, truncated at order eight, to 
the numerical points. In the right panel of Fig.~\ref{fig:calZ}, we have calculated 
$\mathcal{Z}_{A, q}^{(n)}$ for $q=L/2$ as a function of $L$ both diagonalising exactly the correlation matrix and 
using Eq.~\eqref{det_Z_alpha} (dots) as well as integrating numerically the expression
in Eq.~\eqref{fourier_charged_mom_n_0_1} (continuous lines). In the latter case, we have
performed the integration both neglecting the $O(1)$ term (red line) and including it by
taking the function fitted in the left panel of this figure (blue line). From this plot it
is clear that Eq.~\eqref{fourier_charged_mom_n_0_1} matches with the exact numerical points 
if we take into account the $O(1)$ term. 

In Appendix~\ref{app_large_L}, we obtain the large $L$ expansion of  $\mathcal{Z}_{A, q}^{(n)}$ from the integral in
Eq.~\eqref{fourier_charged_mom_n_0_1}. In particular, using Eq.~\eqref{eq:qmoments} in the appendix, 
we conclude that at leading order $\mathcal{Z}_{A, q}^{(n)}$ are Gaussian functions of $q$
\begin{equation}\label{eq:znq}
 \mathcal{Z}_{A,q}^{(n)}\sim Z_{A}^{(n)}(0)\frac{e^{-\frac{(q-L/2)^2}{4\left|\Delta_n^{(2)}(\nu, \delta)\right|\log L}}}{2\sqrt{\pi\left|\Delta_n^{(2)}(\nu,\delta)\right|\log L}},
\end{equation}
centred at $q=L/2$ and with variance $2\Delta_n^{(2)}(\nu, \delta)\log L$.
Recall that $\Delta_n^{(2)}$ is the coefficient of the $\alpha^2$ term of 
the quantity $\Delta_n(\nu, \delta, \alpha)$ that we introduced in Eq.~\eqref{Delta_n_power_series}.
Plugging this result into Eq.~\eqref{symm_res_ent_fourier}, we find that the 
symmetry-resolved R\'enyi entropy in the ground state of the long-range 
SSH model behaves for large $L$ as
\begin{equation}\label{eq:entropies}
 S_{A, q}^{(n)}= S_{A}^{(n)}-\frac{1}{2}\log\log L+
 \Upsilon_n(\nu, \delta)\frac{(q-L/2)^2}{\log L}+\Upsilon_n'(\nu, \delta)+o(1/\log L),
\end{equation}
in the interval $0\leq \nu <1$, with
\begin{equation}\label{Upsilon_n}
 \Upsilon_n(\nu, \delta)=\frac{1}{4(n-1)}
 \left[\frac{1}{\left|\Delta_n^{(2)}(\nu, \delta)\right|}
 -\frac{n}{\left|\Delta_1^{(2)}(\nu,\delta)\right|}\right],
\end{equation}
and 
\begin{equation}\label{Upsilonp_n}
\Upsilon_n'(\nu, \delta)=\frac{1}{1-n}\log\left[(2\sqrt{\pi})^{n-1}
\frac{\left|\Delta_1^{(2)}(\nu, \delta)\right|^{n/2}}
{\left|\Delta_n^{(2)}(\nu, \delta)\right|^{1/2}}\right].
\end{equation}

\begin{figure}[t]
 \begin{minipage}{0.5\linewidth}
 \centering 
 \includegraphics[width=\textwidth]{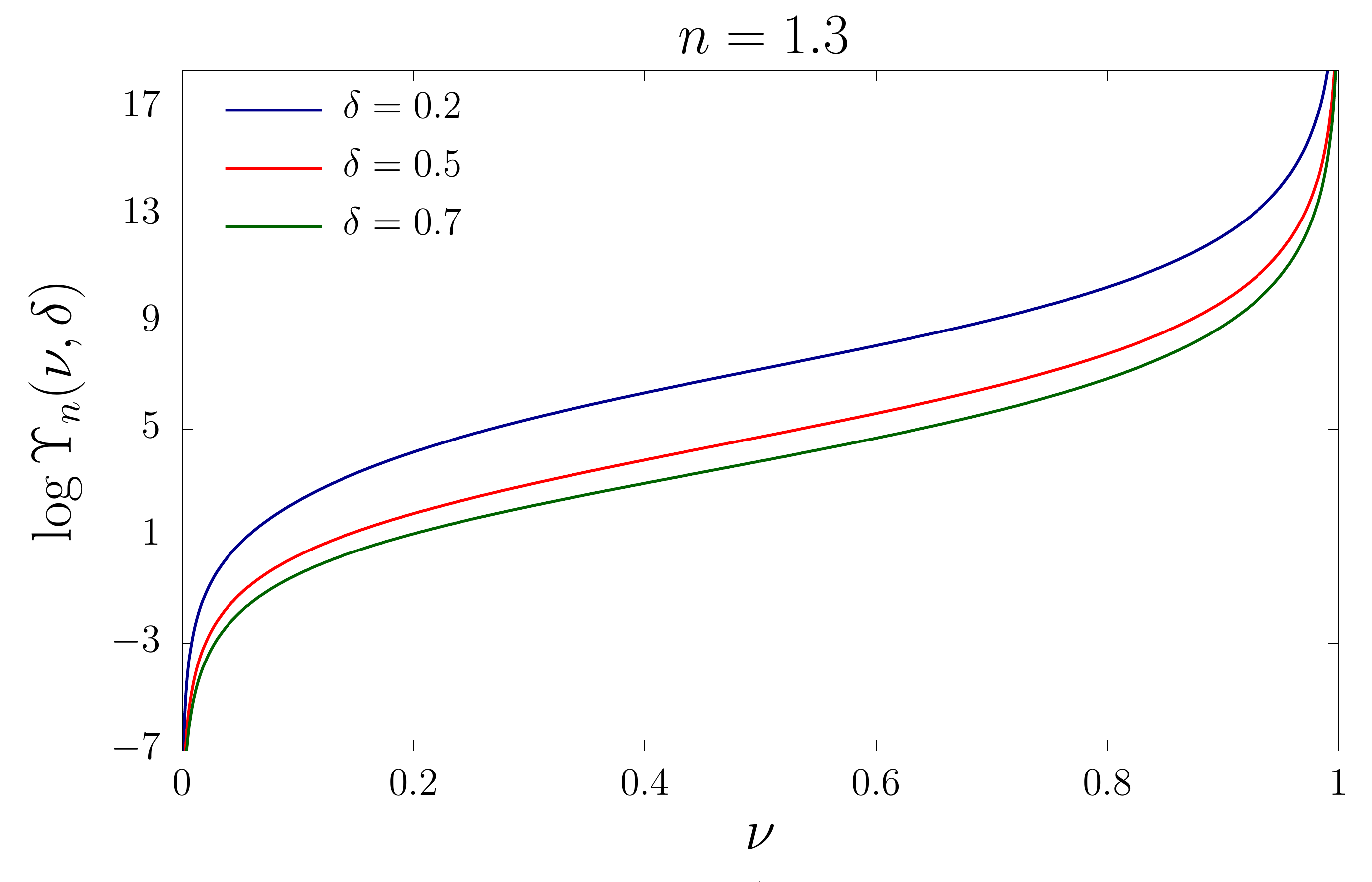}
\end{minipage}
 \begin{minipage}{0.5\linewidth}
 \centering 
 \includegraphics[width=\textwidth]{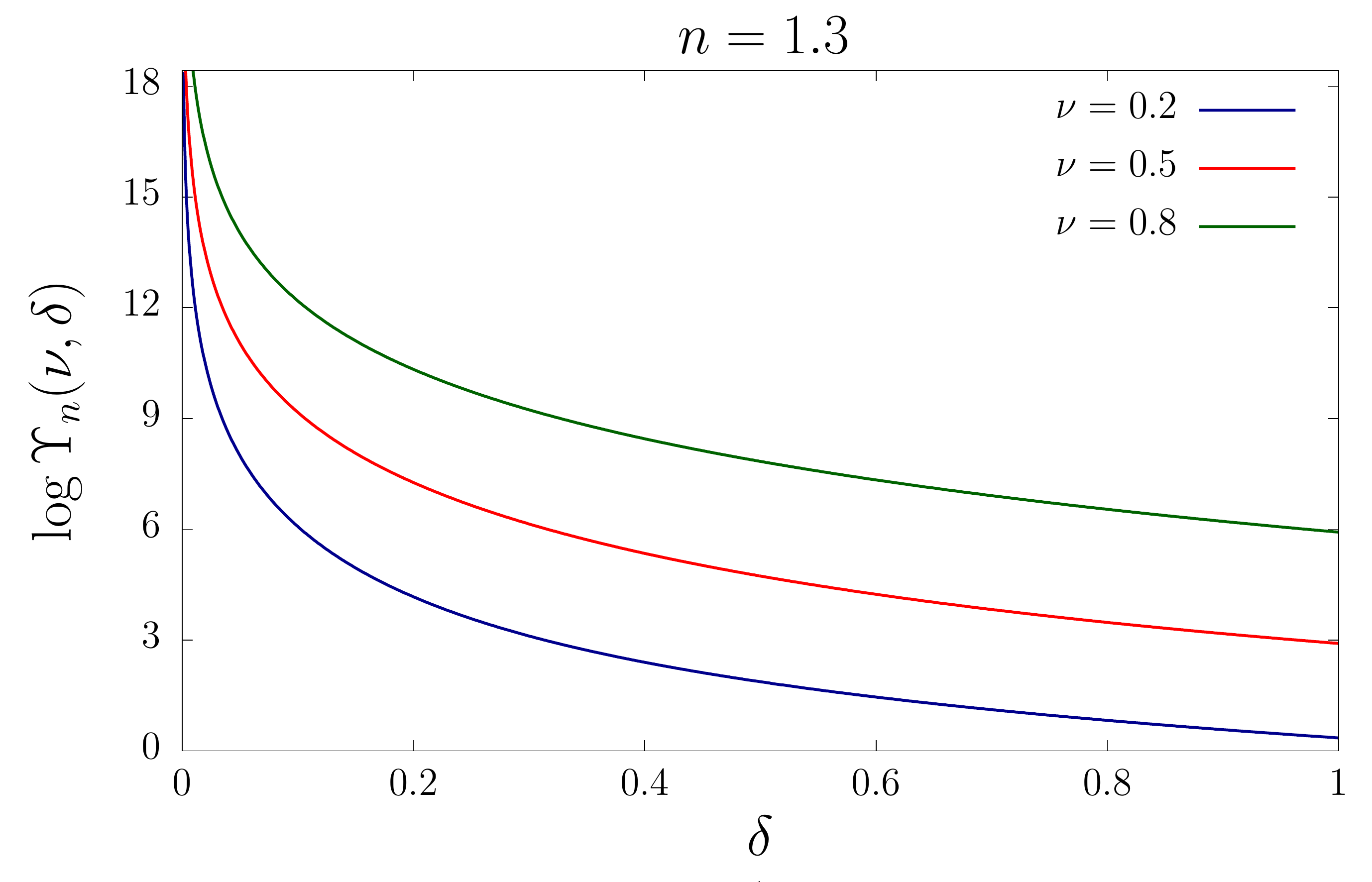}
\end{minipage}
\begin{center}
 \begin{minipage}{0.5\linewidth}
  \includegraphics[width=\textwidth]{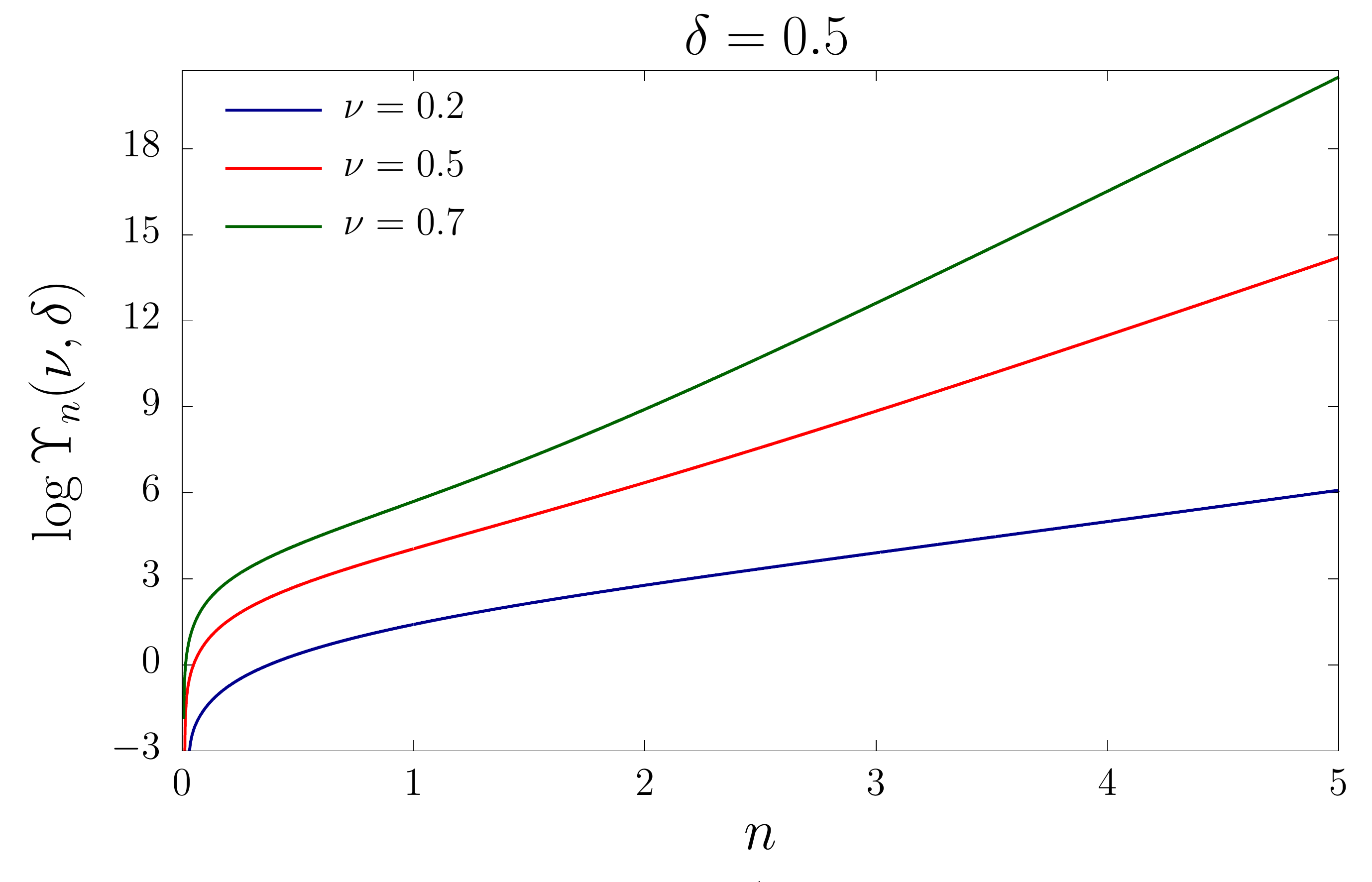}
 \end{minipage}
\end{center}
\caption{Plot of the logarithm of the coefficient $\Upsilon_n(\nu, \delta)$ given in Eq.~\eqref{Upsilon_n} as 
         a function of $\nu$ (upper left panel), $\delta$ (upper right panel) and $n$ (lower 
         panel).}\label{fig:Upsilon_n}
\end{figure}

\begin{figure}[t]
 \begin{minipage}{0.5\linewidth}
 \centering 
 \includegraphics[width=\textwidth]{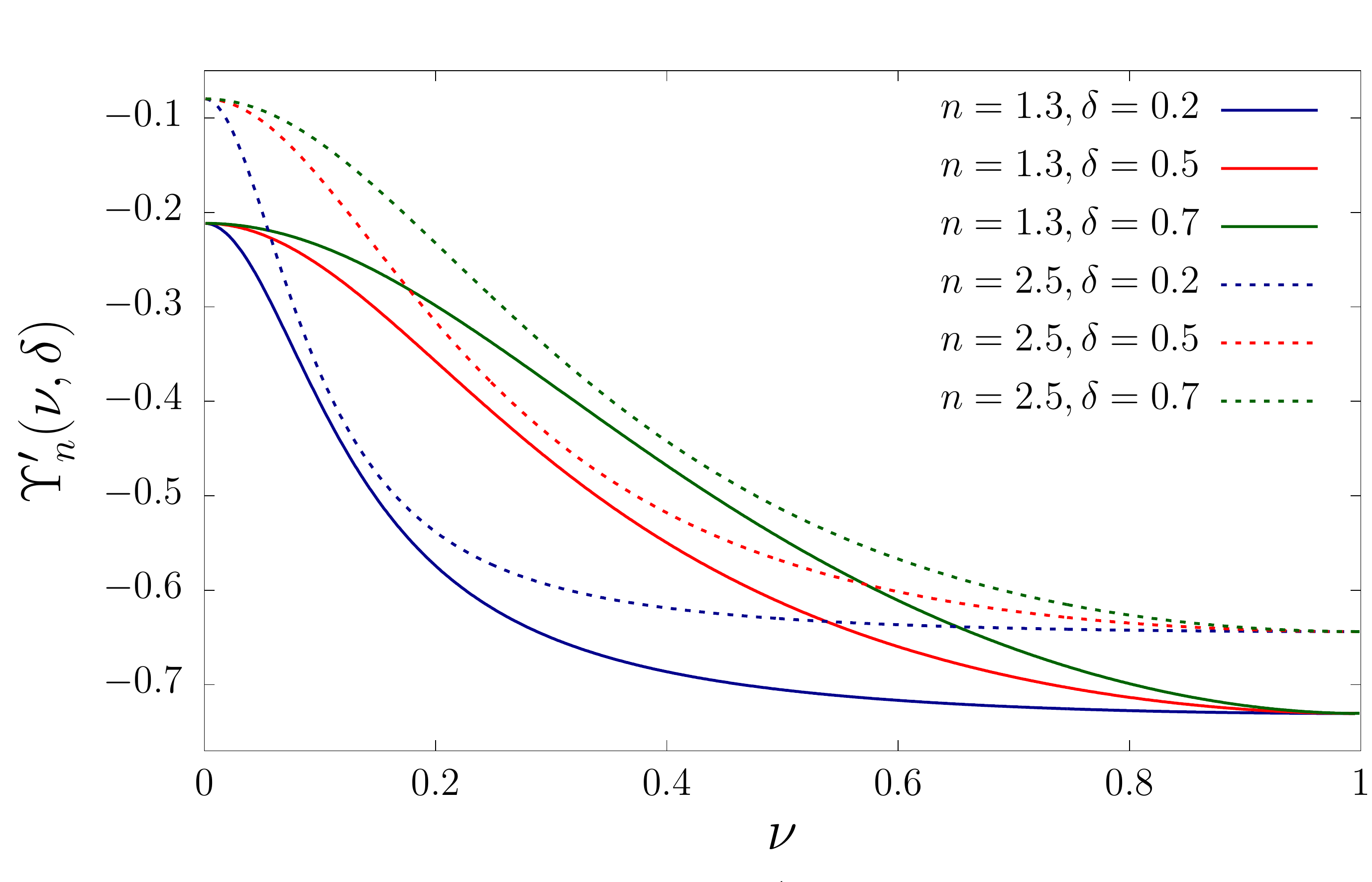}
\end{minipage}
 \begin{minipage}{0.5\linewidth}
 \centering 
 \includegraphics[width=\textwidth]{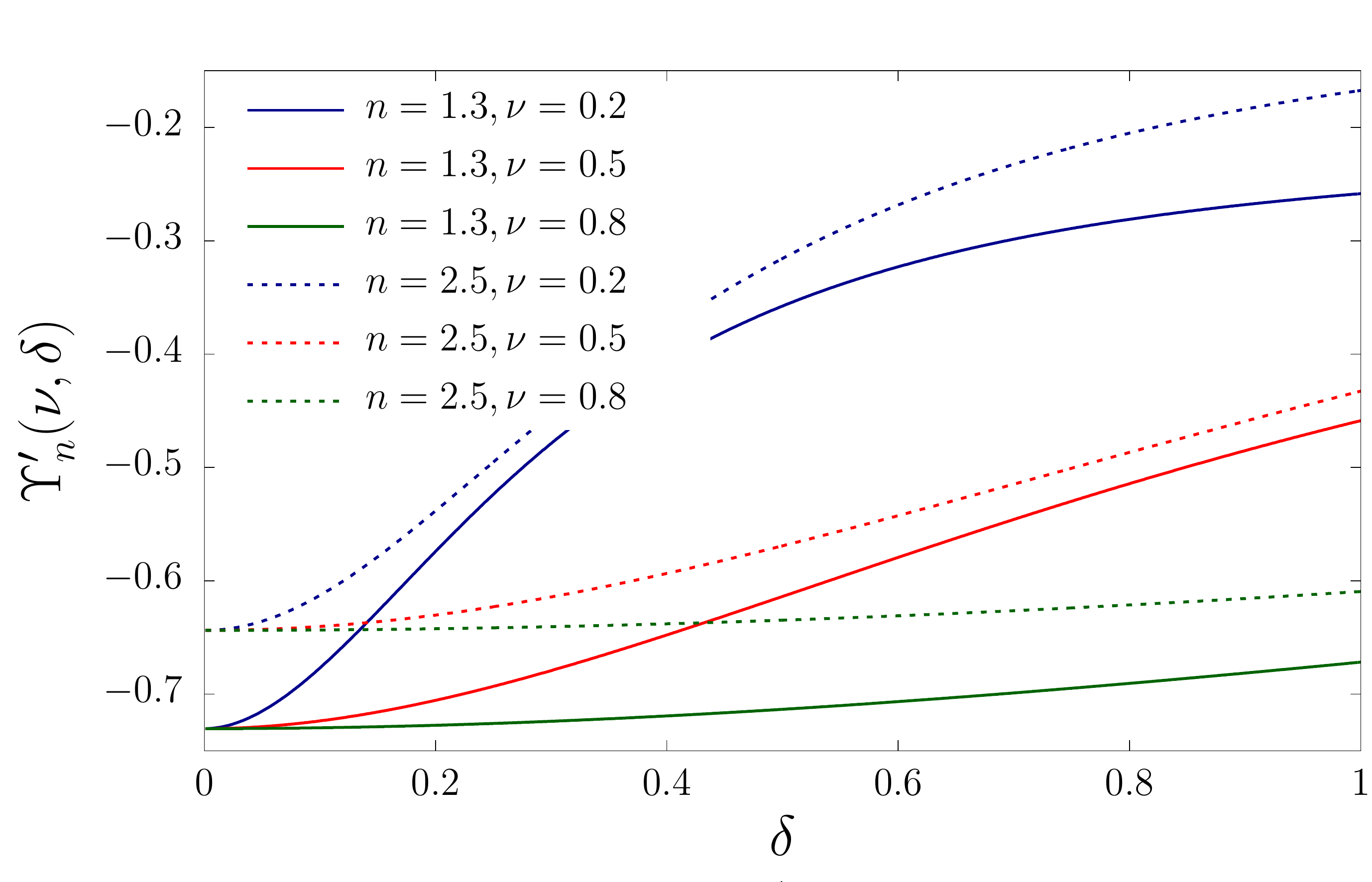}
\end{minipage}
\begin{center}
 \begin{minipage}{0.5\linewidth}
  \includegraphics[width=\textwidth]{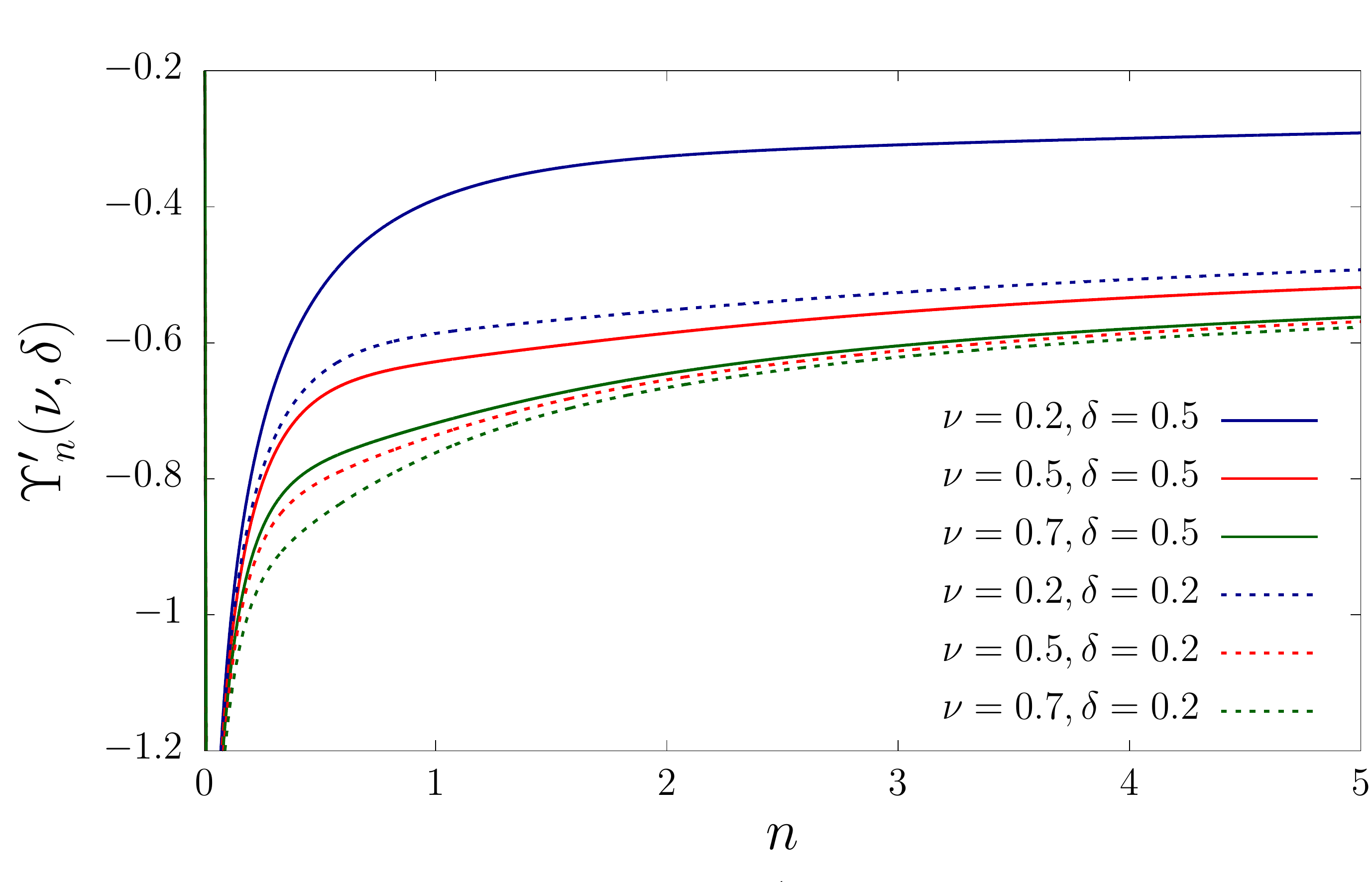}
 \end{minipage}
\end{center}
\caption{Plot of the coefficient $\Upsilon_n'(\nu, \delta)$ given in Eq.~\eqref{Upsilonp_n} as 
         a function of $\nu$ (upper left panel), $\delta$ (upper right panel) and $n$ (lower 
         panel).}\label{fig:Upsilonp_n}
\end{figure}

Let us discuss the result in Eq.~\eqref{eq:entropies}. The leading terms for large $L$ (up to $O(1)$) 
do not depend on the charge $q$; at first order in $L$, they are given by the total R\'enyi entanglement entropy $S_A^{(n)}$ in Eq.~\eqref{full_renyi_ee_ssh}. 
Moreover, note that the coefficient of the double logarithmic correction is $1/2$, as in the short-range case.
In Ref.~\cite{nonabelian-21}, it was proven that, for CFTs with an internal Lie group symmetry, the coefficient of such term is
equal to half of the dimension of the group, which here is $U(1)$. Thus, somehow, the long-range hoppings do not 
spoil this result. The first term that breaks equipartition is at order $O((\log L)^{-1})$ 
and its amplitude is governed by the coefficient $\Upsilon_n(\nu, \delta)$ defined in Eq.~\eqref{Upsilon_n}
and plotted in Fig.~\ref{fig:Upsilon_n}. The presence of such 
term is a novelty with respect to what happens in the critical short-range systems, where the first term breaking 
the equipartition occurs at order $O((\log L)^{-2})$. As we saw in Section~\ref{sec:definitions}, in the case of 
the tight-binding model, the modes $\mathcal{Z}_{A, q}^{(n)}$ are also Gaussian in the large $L$ limit, but the variance is proportional to $1/n$, cf. Eq.~\eqref{eq:ressaddleU1}. The latter implies that 
the term of order $O((\log L)^{-1})$ in the symmetry-resolved entropy vanishes. On the contrary, in our case, the variance,
which is given by the coefficient $\Delta_n^{(2)}(\nu, \delta)$, has a complicated dependence on $n$.  
For this reason, the $O((\log L)^{-1})$ term does not cancel in the long-range case, as it would occur if 
$\Delta_n^{(2)}(\nu, \delta) \propto 1/n$. This is precisely the situation at $\nu=0$, for which $\Delta_n^{(2)}(0, \delta)=-1/(2\pi^2 n)$, 
and therefore $\Upsilon_n(\nu=0,\delta)=0$, i.e. the correction at $O((\log L)^{-1})$ disappears. In this case, the 
additive term $\Upsilon_n'(\nu,\delta)$ in Eq.~\eqref{eq:entropies} reads
\begin{equation}
\Upsilon_n'(\nu=0, \delta)=\frac{1}{2}\frac{\log n}{1-n}+\frac{1}{2}\log \frac{\pi}{2},
\end{equation}
and it is the same as that of the tight-binding model, see Eq.~\eqref{eq:intro}.
In Fig.~\ref{fig:Upsilonp_n}, we plot $\Upsilon_n'(\nu, \delta)$ as a function of $\nu$, $\delta$ and $n$.

\begin{figure}[t]
 \begin{minipage}{0.5\linewidth}
 \centering 
 \includegraphics[width=\textwidth]{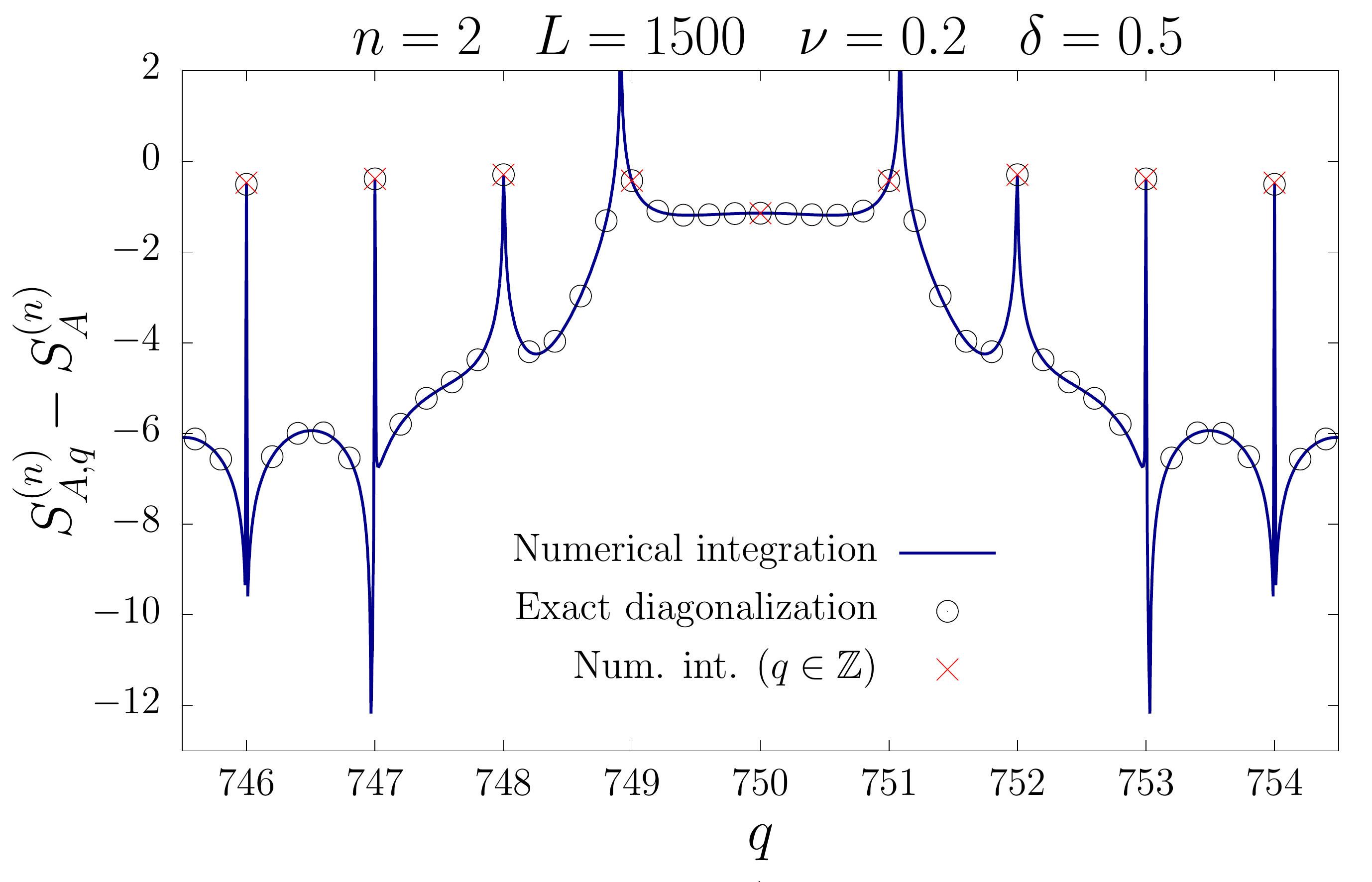}
\end{minipage}
 \begin{minipage}{0.5\linewidth}
 \centering 
 \includegraphics[width=\textwidth]{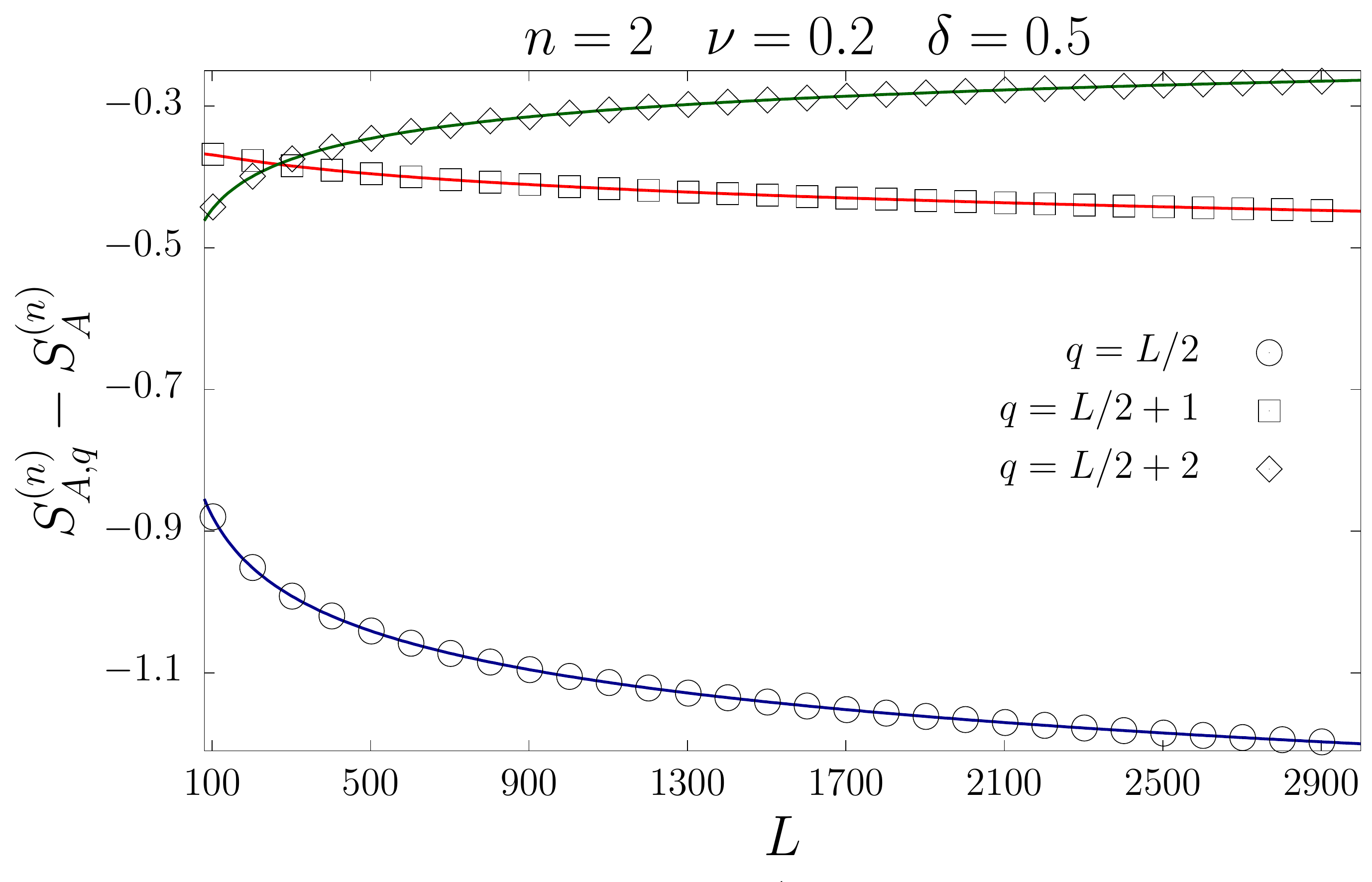}
\end{minipage}
\caption{In the left panel, we study the symmetry-resolved $n=2$ R\'enyi entropy 
as a function of the charge $q$ for an interval of fixed length $L=1500$. Note that we subtract 
the entanglement entropy $S_A^{(n)}$. The points $\circ$
have been obtained by computing numerically the charged moments through Eq.~\eqref{det_Z_alpha}. The continuous 
line corresponds to integrating numerically Eq.~\eqref{fourier_charged_mom_n_0_1}, taking as $O(1)$ term the function fitted 
in the left plot of Fig.~\ref{fig:calZ}, $\sum_{j=1}^8 b_{2j}\alpha^{2j}$. We indicate by $\textcolor{red}{\times}$
the values that this curve takes when $q$ is an integer.
In the right panel, we analyse the symmetry-resolved $n=2$ R\'enyi entropy by varying the length of the interval $L$
for different fixed values of the charge. As in the left plot, the points have been obtained by calculating
numerically the charged moments applying Eq.~\eqref{det_Z_alpha} while the lines correspond to integrating numerically Eq.~\eqref{fourier_charged_mom_n_0_1} with the 
same $O(1)$ term as before.}\label{fig:symm_res_ent}
\end{figure}

\begin{figure}[t]
 \begin{minipage}{0.5\linewidth}
 \centering 
 \includegraphics[width=\textwidth]{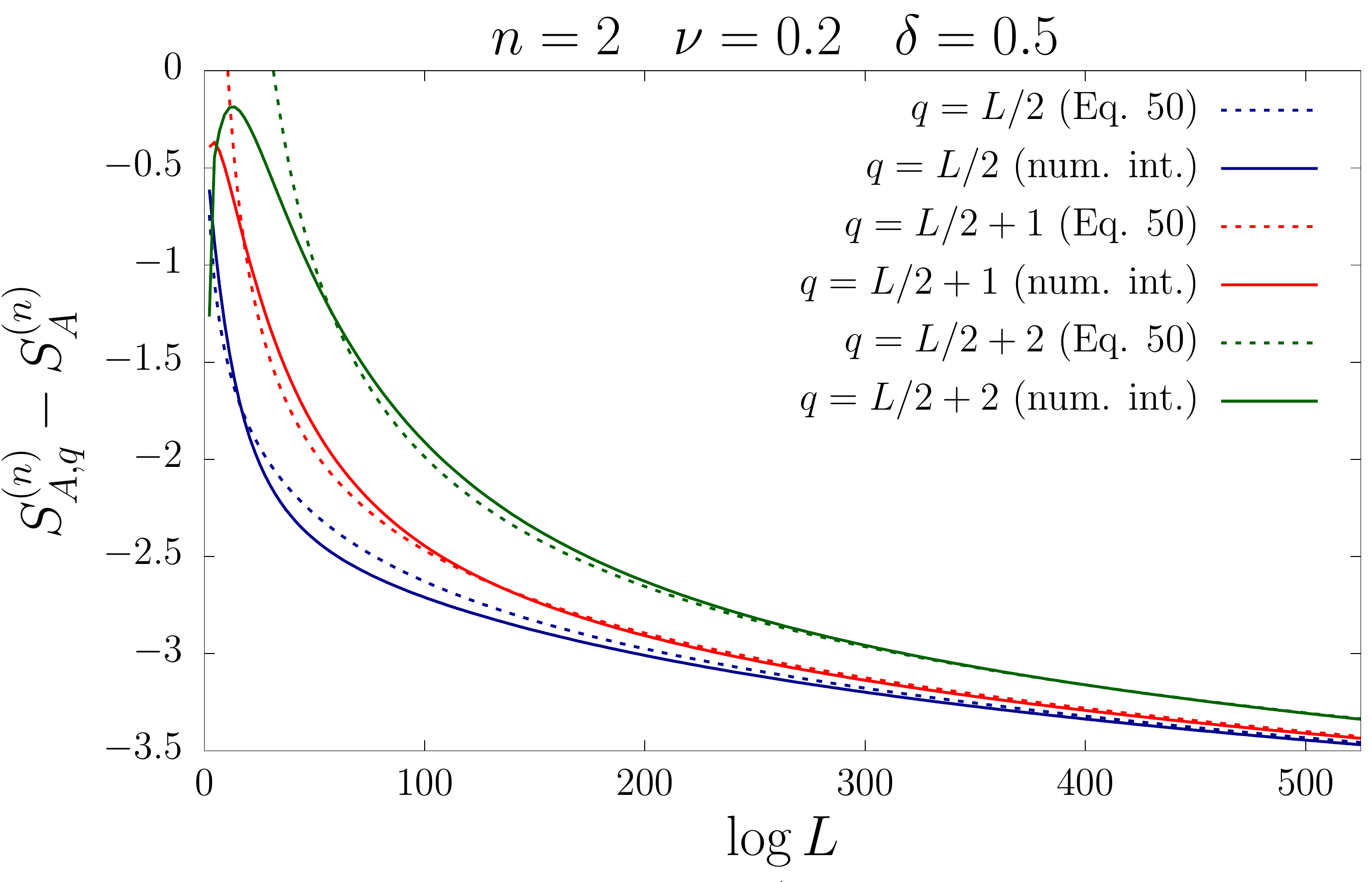}
\end{minipage}
 \begin{minipage}{0.5\linewidth}
  \centering
  \includegraphics[width=\textwidth]{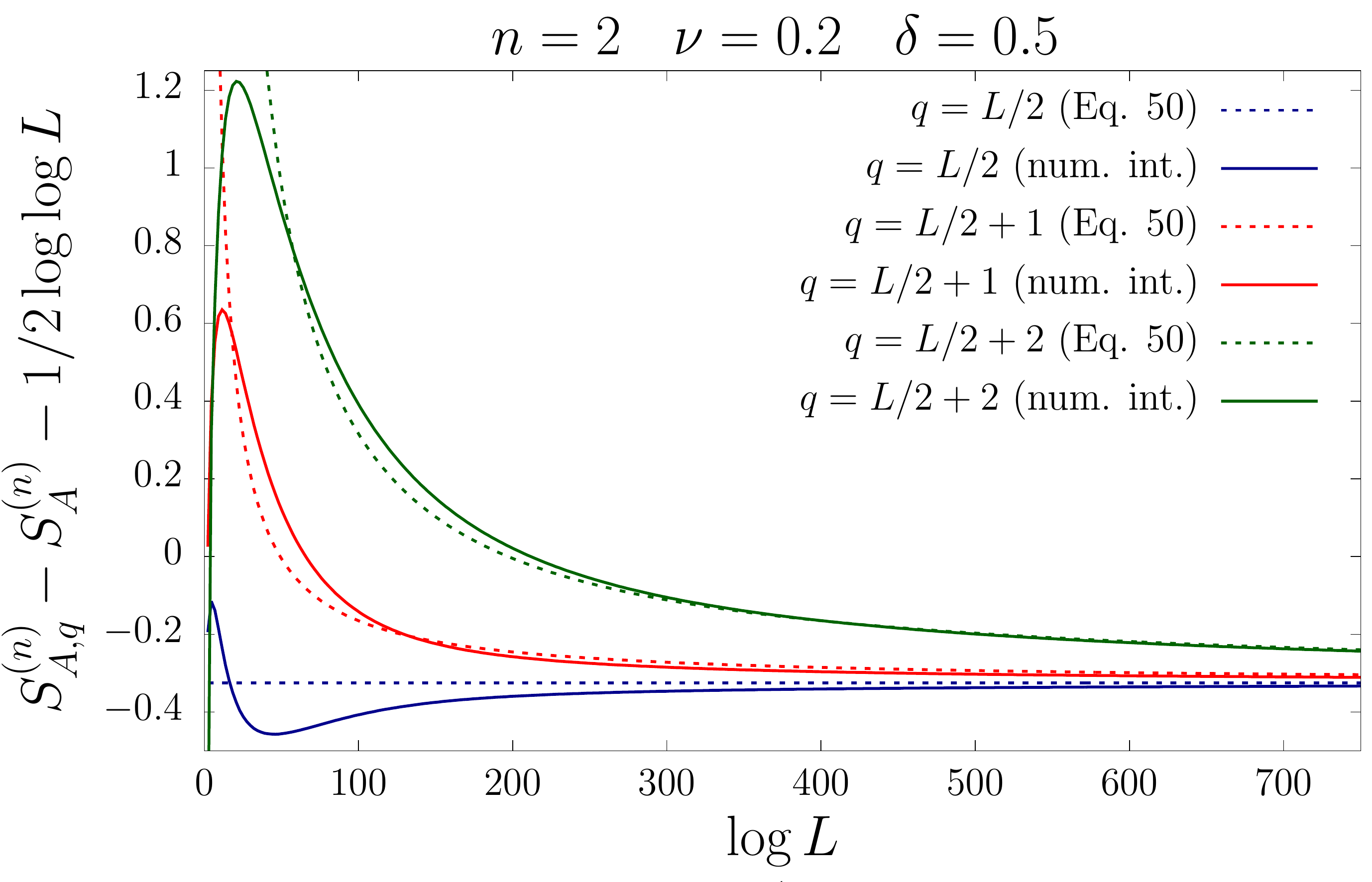}
 \end{minipage}
\caption{Analysis of the symmetry-resolved $n=2$ R\'enyi entropy as a function of the length of the interval $L$ for different 
charges $q$. The continuous curves represent the result obtained by integrating numerically Eq.~
\eqref{fourier_charged_mom_n_0_1}, using the function fitted in the left panel of Fig.~\ref{fig:calZ}
as $O(1)$ term, while the dashed ones correspond to the asympotic prediction of 
Eq.~\eqref{eq:entropies}. Note that, comparing with Fig.~\ref{fig:symm_res_ent}, here we 
have considered larger intervals, up to $L=10^{500}$. In the plot on the left, we have removed 
the contribution from the R\'enyi entanglement entropy while in the right
one we have also subtracted the term $-1/2\log\log L$.}\label{fig:symm_res_ent_large_L}
\end{figure}

The asymptotic expression found in Eq.~\eqref{eq:entropies} properly describes the 
symmetry-resolved entanglement entropies for values of $L$ that are not accessible 
numerically by diagonalising the correlation matrix. In Fig.~\ref{fig:symm_res_ent}, 
we check that the symmetry-resolved entropies obtained by integrating numerically 
Eq.~\eqref{fourier_charged_mom_n_0_1} match the exact values computed from the 
diagonalization of the correlations. Nevertheless, for the range of $L$ considered in 
the right panel of this figure, the symmetry-resolved entropies are still far from behaving as Eq.~\eqref{eq:entropies} 
predicts in the large $L$ limit. In fact, from that plot is clear that the difference 
$S_{A, q}^{(n)}-S_{A}^{(n)}$ is not in general a monotonic decreasing function on $L$, 
as Eq.~\eqref{eq:entropies} actually is. In Fig.~\ref{fig:symm_res_ent_large_L}, we take 
larger values of $L$ and we show that the symmetry-resolved entropies computed through 
the numerical integration of Eq.~\eqref{fourier_charged_mom_n_0_1} tend to the asymptotic 
expression of Eq.~\eqref{eq:entropies} when $L>10^{200}$, an unreasonably large number. 
Such a gigantic scale is not a consequence of any peculiar feature of the long-range SSH model, 
but just a very unfortunate coincidence. In general, one does not expect such enormous 
scale to observe the predicted asympotic behaviour. For example, in the tight-binding model, 
the symmetry-resolved entanglement entropy tends to the asymptotic behaviour reviewed in Sec.~\ref{sec:definitions} 
when the length of the interval is $L>10^2$, see Ref.~\cite{riccarda}.

\section{Conclusions}\label{sec:concl}

In this manuscript, we have derived exact results for the symmetry-resolved entanglement entropies
of an interval of size $L$ in an infinite quadratic fermionic chain with dimerised long-range couplings, decaying as a power law with the distance. 
A general feature of systems with this kind of couplings is that they effectively behave as a short-range model above a certain 
value of the dumping exponent $\nu$ of the coupling, while below such value they genuinely display a 
long-range character~\cite{Defenu}. Here we have seen that this also occurs when we study the symmetry resolution of 
the ground state entanglement. We have first investigated the charged moments of the reduced density 
matrix. When $\nu\geq 1$, they satisfy an area law as in gapped short-range chains. On the contrary, for 
$0\leq \nu<1$, there is a logarithmic correction despite the system is gapped too. The prefactor of this 
logarithmic term is a non-universal function which does explicitly  depend on the long-range couplings
(see Eqs.~\eqref{Z_alpha_n_0_1}~and~\eqref{eq:B_n1}).
We arrived to this result by exploiting the relation between the two-point correlation functions
and charged moments, as well as the block Toeplitz structure of the former. In particular, the presence
of discontinuities in the entries of the correlation matrix when $0\leq \nu <1$ leads to the 
logarithmic growth of the charged moments in that regime. 

We then focused on the symmetry-resolved entropies, which can be calculated from the Fourier transform of the charged moments. 
We have found an analytical expansion for them (see Eq.~\eqref{eq:entropies}) at large $L$ in the long-range regime $0\leq \nu<1$. 
An important aspect in the analysis of these quantities is their dependence on the symmetry sector. We have determined the 
first term in the large $L$ expansion that depends on the charge. The novelty with respect to the results concerning 
the tight-binding model is that the entanglement equipartition is broken at $O((\log L)^{-1})$, rather than $O((\log L)^{-2})$.
Moreover, the coefficient  of this term varies with $\nu$, and vanishes when $\nu=0$.

Let us conclude this manuscript with some possible directions for future investigations. 
Here we have taken as subsystem a single interval of contiguous sites. Our analysis 
can be straightforwardly  extended to disjoint intervals by applying the results on
minors of (block) Toeplitz matrices derived in Ref.~\cite{Ares3} for the study of the 
entanglement entropy of disjoint intervals in quadratic, homogeneous fermionic chains. 
This multipartite geometry also opens the way to examine the entanglement negativity, which 
is a quantifier of the entanglement in mixed states both total \cite{cct-12,cct-13,ssr-17,sr-19,ryu,mvdc-22} and symmetry-resolved~\cite{goldstein1, mbc-21, cn-21,vecd1-20}. Finally, 
it would be interesting to explore the effect of the long-range couplings also in interacting 
systems, e.g. the XXZ spin chain~\cite{Defenu}: one may wonder whether the universal prefactor 
of the logarithmic term of the charged moments \cite{Goldstein} keeps its universal behaviour 
or is affected by the long-range couplings, as happens in the (dimerised) free case.
A final intriguing related calculation concerns the study of the symmetry resolution of long-range hierarchical models,  
generalising known results for the total entanglement \cite{pcp-19,m-20}.

\textit{Note added.---} After the completion of this work, Ref.~\cite{Jones} appeared in arXiv, where
symmetry-resolved entanglement entropy in critical free-fermionic chains with long-range hoppings is also considered.

\section*{Acknowledgments}

The authors acknowledge support from ERC under Consolidator grant number 771536 (NEMO).

\begin{appendices}

\section*{Appendix}

\section{Large $L$ behaviour of Fourier modes $\mathcal{Z}_{A, q}^{(n)}$}\label{app_large_L}

In this Appendix, we will determine the large $L$ behaviour of modes $\mathcal{Z}_{A, q}^{(n)}$ 
from the integral in Eq.~\eqref{fourier_charged_mom_n_0_1}. To this end, let us rewrite the 
coefficient $\Delta_n(\nu, \delta, \alpha)$ that appears in the exponent of the integrand 
in the power series of Eq.~\eqref{Delta_n_power_series},
\begin{equation}
 \mathcal{Z}_{A, q}^{(n)}=\frac{Z_{A}^{(n)}(0)}{2\pi}\int_{-\pi}^\pi
 e^{-i\alpha(q-L/2)}e^{\sum_{j=1}^\infty \Delta_{2j}\alpha^{2j}\log L}d\alpha,
\end{equation}
where we introduced the shorthand notation $\Delta_{2j}\equiv\Delta_n^{(2j)}(\nu, \delta, \alpha)$.
If we now perform the change of variables 
\begin{equation}
 \tilde{\alpha}=\alpha\sqrt{|\Delta_2|\log L},
\end{equation}
then we have 
\begin{equation}
 \mathcal{Z}_{A, q}^{(n)}\sim \frac{Z_{A}^{(n)}(0)}{2\pi\sqrt{|\Delta_2|\log L}}
 \int_{-\pi\sqrt{|\Delta_2|\log L}}^{\pi\sqrt{|\Delta_2|\log L}}
 e^{-\frac{i\tilde{\alpha}(q-L/2)}{\sqrt{|\Delta_2|\log L}}}
 e^{-\tilde{\alpha}^2+\sum_{j=2}^\infty\frac{\Delta_{2j}}{|\Delta_2|^j(\log L)^{j-1}}\tilde{\alpha}^{2j}}
 d\tilde{\alpha}.
\end{equation}
Now we truncate at order $2M$ the series in the exponent of the integrand and 
we expand in Taylor series each exponential $e^{\frac{\Delta_{2j}}{|\Delta_2|^j(\log L)^{j-1}}\tilde{\alpha}^{2j}}$.
Then we obtain
\begin{multline}
 \mathcal{Z}_{A, q}^{(n)}\sim \frac{Z_{A}^{(n)}(0)}{2\pi\sqrt{|\Delta_2|\log L}}
 \sum_{k_1,\dots, k_{M-1}=0}^\infty \frac{\Delta_4^{k_1}\Delta_6^{k_2}\cdots
 \Delta_{2M}^{k_{M-1}}}{|\Delta_2|^{2k_1+3k_2+\cdots+Mk_{M-1}}}\frac{1}{(\log L)^{k_1+2k_2+\cdots+(M-1)k_{M-1}}}\times \\
 \frac{1}{k_1!k_2!\cdots k_{M-1}!}\int_{-\pi\sqrt{|\Delta_2|\log L}}^{\pi\sqrt{|\Delta_2|\log L}}
 e^{-\frac{i\tilde{\alpha}(q-L/2)}{\sqrt{|\Delta_2|\log L}}}e^{-\tilde{\alpha}^2}\tilde{\alpha}^{4k_1+6k_2+\cdots+2Mk_{M-1}}d\tilde{\alpha}.
\end{multline}
In the large $L$ limit, the integrals of the previous expression tend to 
\begin{equation}
\begin{split}
 \int_{-\pi\sqrt{|\Delta_2|\log L}}^{\pi\sqrt{|\Delta_2|\log L}}
 e^{-\frac{i\tilde{\alpha}(q-L/2)}{\sqrt{|\Delta_2|\log L}}}e^{-\tilde{\alpha}^2}\tilde{\alpha}^{2k}d\tilde{\alpha}=&
 \int_{-\infty}^{\infty}
 e^{-\frac{i\tilde{\alpha}(q-L/2)}{\sqrt{|\Delta_2|\log L}}}e^{-\tilde{\alpha}^2}\tilde{\alpha}^{2k}d\tilde{\alpha}
 +O(e^{-|\Delta_2|\pi^2\log L}(\log L)^k)\\
 =& \Gamma\left(\frac{1}{2}+k\right) {_1}F_1\left(\frac{1}{2}+k, \frac{1}{2}, -\frac{(q-L/2)^2}{4|\Delta_2|\log L}\right) \\
 & +O(e^{-|\Delta_2|\pi^2\log L}(\log L)^k) .
\end{split}
\end{equation}
Thus
\begin{multline}
 \mathcal{Z}_{A, q}^{(n)}\sim \frac{Z_{A}^{(n)}(0)}{2\pi\sqrt{|\Delta_2|\log L}}
 \sum_{k_1,\dots, k_{M-1}=0}^\infty \frac{\Delta_4^{k_1}\cdots
 \Delta_{2M}^{k_{M-1}}}{|\Delta_2|^{2k_1+\cdots+Mk_{M-1}}}\frac{\Gamma(1/2+2k_1+\cdots+M k_{M-1})}{(\log L)^{k_1+\cdots+(M-1)k_{M-1}}}\times \\
 \frac{1}{k_1!\cdots k_{M-1}!} {_1}F_1\left(\frac{1}{2}+2k_1+\cdots+Mk_{M-1}, \frac{1}{2}, -\frac{(q-L/2)^2}{4|\Delta_2|\log L}\right) .
\end{multline}
If we apply the Kummer transformation of the confluent hypergeometric function~\cite{NIST}, 
we finally find
\begin{multline}\label{eq:qmoments}
 \mathcal{Z}_{A,q}^{(n)}\sim Z_{A}^{(n)}(0)\frac{e^{-\frac{(q-L/2)^2}{4|\Delta_2|\log L}}}{2\pi\sqrt{|\Delta_2|\log L}}
 \sum_{k_1,\dots,k_{M-1}=0}^{\infty} \frac{\Delta_4^{k_1}\cdots
 \Delta_{2M}^{k_{M-1}}}{|\Delta_2|^{2k_1+\cdots+Mk_{M-1}}}\frac{\Gamma(1/2+2k_1+\cdots+M k_{M-1})}{(\log L)^{k_1+\cdots+(M-1)k_{M-1}}}\times \\
 \frac{1}{k_1!\cdots k_{M-1}!} {_1}F_1\left(-2k_1-\cdots-Mk_{M-1}, \frac{1}{2}, \frac{(q-L/2)^2}{4|\Delta_2|\log L}\right). 
\end{multline}
At leading order in $L$, Eq. \eqref{eq:qmoments} reduces to Eq. \eqref{eq:znq} in the main text.
\end{appendices}


\section*{References}
\begin{thebibliography}{10}


\bibitem{Bombelli} 
L. Bombelli, R. K. Koul, J. Lee, and R. Sorkin, 
\textit{Quantum source of entropy for black holes}, 
\href{https://doi.org/10.1103/PhysRevD.34.373}{Phys. Rev. D {\bf 34} 373 (1986)}.
 
\bibitem{srednicki}
M. Srednicki, 
{\it Entropy and area},
\href{https://link.aps.org/doi/10.1103/PhysRevLett.71.666}{Phys. Rev. Lett. {\bf 71}, 5 (1993)}.

\bibitem{dong}
X.~Dong,
{\it The gravity dual of R{\'{e}}nyi entropy},
\href{http://dx.doi.org/10.1038/ncomms12472}{Nature Comm. {\bf 7}, 1 (2016)}.

\bibitem{RT}
S. Ryu and T. Takayanagi, 
{\it Holographic Derivation of Entanglement Entropy from the anti--de Sitter Space/Conformal Field Theory Correspondence},
\href{https://link.aps.org/doi/10.1103/PhysRevLett.96.181602}{Phys. Rev. Lett. {\bf 96}, 18 (2006)}.

\bibitem{Raamsdonk}
M. Van Raamsdonk, 
{\it Building up spacetime with quantum entanglement},
\href{https://doi.org/10.1007/s10714-010-1034-0}{Gen. Rel. and Grav. {\bf 42}, 10 (2010)}.

\bibitem{maldacena}
J. Maldacena and L. Susskind, 
{\it Cool horizons for entangled black holes},
\href{http://dx.doi.org/10.1002/prop.201300020}{Fortschr. Phys. {\bf 61}, 781(2013)}.

\bibitem{h-06}
M. B. Hastings, 
{\it An area law for one-dimensional quantum systems},
\href{http://dx.doi.org/10.1088/1742-5468/2007/08/P08024}{J. Stat. Mech. (2007) P08024}.

\bibitem{cc-04}
P. Calabrese and J. Cardy, 
{\it Entanglement entropy and quantum field theory},
\href{http://dx.doi.org/10.1088/1742-5468/2004/06/P06002}{J. Stat. Mech. (2004) P06002}.

\bibitem{cc-09}
P. Calabrese and J. Cardy, 
{\it Entanglement entropy and conformal field theory},
\href{http://dx.doi.org/10.1088/1751-8113/42/50/504005}{J. Phys. A {\bf 42}, 504005 (2009)}.

\bibitem{hlw-94}
C. Holzhey, F. Larsen, and F. Wilczek, 
{\it Geometric and renormalized entropy in conformal field theory},
\href{http://dx.doi.org/10.1016/0550-3213(94)90402-2}{Nucl. Phys. B {\bf 424}, 443 (1994)}.

\bibitem{vidal}
G. Vidal, J. I. Latorre, E. Rico, and A. Kitaev, 
{\it Entanglement in quantum critical phenomena},
\href{http://dx.doi.org/10.1103/PhysRevLett.90.227902}{Phys. Rev. Lett. {\bf 90}, 227902 (2003)}.

\bibitem{vidal1}
J. I. Latorre, E. Rico, and G. Vidal,
{\it Ground state entanglement in quantum spin chains},
\href{https://arxiv.org/abs/quant-ph/0304098}{Quant. Inf. Comp. {\bf 4}, 048 (2004)}. 

\bibitem{vidal2}
J. I. Latorre and A. Riera, 
{\it A short review on entanglement in quantum spin systems}, 
\href{http://dx.doi.org/10.1088/1751-8113/42/50/504002}{J. Phys. A {\bf 42}, 504002 (2009)}.

\bibitem{Amico}
L.~Amico, R.~Fazio, A.~Osterloh, and V.~Vedral,
\emph{Entanglement in many-body systems},
\href{http://dx.doi.org/10.1103/RevModPhys.80.517}{Rev. Mod. Phys. {\bf 80}, 517 (2008)}.

\bibitem{Laflorencie}
N.~Laflorencie,
{\it Quantum entanglement in condensed matter systems},
\href{http://dx.doi.org/10.1016/j.physrep.2016.06.008}{Phys. Rep. {\bf 643}, 1 (2016)}.

\bibitem{dlegp-14}
D. Vodola, L. Lepori, E. Ercolessi, A. V. Gorshkov, and G. Pupillo, 
{\it Kitaev chains with longrange pairing}, 
\href{http://dx.doi.org/10.1103/PhysRevLett.113.156402}{Phys. Rev. Lett. {\bf 113}, 156402 (2014)}.

\bibitem{Ares0} 
F. Ares, J. G. Esteve, F. Falceto, and A. R. de Queiroz, 
\textit{Entanglement in fermionic chains with finite-range coupling and broken symmetries}, 
\href{https://doi.org/10.1103/PhysRevA.92.042334}{Phys. Rev. A {\bf 92}, 042334 (2015)}.

\bibitem{dlegp-16}
D. Vodola, L. Lepori, E. Ercolessi, A. V. Gorshkov, and G. Pupillo, 
{\it Long-range Ising and Kitaev models: phases, correlations and edge modes}, 
\href{http://dx.doi.org/10.1088/1367-2630/18/1/015001}{New J. Phys. {\bf 18}, 015001 (2016)}.

\bibitem{Ares1} 
F. Ares, J. G. Esteve, F. Falceto, and A. R. de Queiroz, 
\textit{Entanglement entropy in the long-range Kitaev chain}, 
\href{https://doi.org/10.1103/PhysRevA.97.062301}{Phys. Rev. A {\bf 97}, 062301 (2018)}.
 
\bibitem{Ares2} 
F. Ares, J. G. Esteve, F. Falceto, and Z. Zimborás, 
\textit{Sublogarithmic behaviour of the entanglement entropy in fermionic chains}, 
\href{http://dx.doi.org10.1088/1742-5468/ab38b6}{J. Stat. Mech. (2019) 093105}.

\bibitem{Defenu} 
N. Defenu, T. Donner, T. Macri, G. Pagano, S. Ruffo, and A. Trombettoni, 
\textit{Long-range interacting quantum systems},
\href{https://arxiv.org/abs/2109.01063}{arXiv:2109.01063 [cond-mat.quant-gas]}.

\bibitem{Saffman} 
M. Saffman, T. G. Walker, and K. Mølmer, 
\textit{Quantum information with Rydberg atoms},
\href{https://doi.org/10.1103/RevModPhys.82.2313}{Rev. Mod. Phys. {\bf 82}, 2313 (2010)}.

\bibitem{Monroe} 
C. Monroe, W. C. Campbell, L.-M. Duan, Z.-X. Gong, A. V. Gorshkov, 
P. W. Hess, R. Islam, K. Kim, N. M. Linke, G. Pagano, P. Richerme, C. Senko, and N. Y. Yao, 
\textit{Programmable quantum simulations of spin systems with trapped ions},
\href{https://doi.org/10.1103/RevModPhys.93.025001}{Rev. Mod. Phys. {\bf 93}, 025001 (2021)}.

\bibitem{Mivehar} F. Mivehvar, F. Piazza, T. Donner, and H. Ritsch, 
\textit{Cavity QED with Quantum Gases: New Paradigms in Many-Body Physics},
\href{https://doi.org/10.1080/00018732.2021.1969727}{Advances in Physics {\bf 70}, 1 (2021)}.

\bibitem{brydges-2018}
T.~Brydges, A. Elben, P. Jurcevic, B.~Vermersch, C. Maier, B. P. Lanyon,  P. Zoller, R. Blatt, and C. F. Roos,
  \textit{Probing entanglement entropy via randomized measurements},
 \href{http://dx.doi.org/10.1126/science.aau4963}{Science {\bf 364}, 260 (2019)}.
 
 
 \bibitem{ekh-20}
A. Elben, R. Kueng, H.-Y. Huang, R. van Bijnen, C. Kokail, M. Dalmonte, P. Calabrese, B. Kraus, J. Preskill, P. Zoller, and B. Vermersch,
{\it Mixed-state entanglement from local randomized measurements}, 
\href{https://doi.org/10.1103/PhysRevLett.125.200501}{Phys. Rev. Lett. {\bf 125}, 200501 (2020)}.

\bibitem{vecd-20}
V. Vitale, A. Elben, R. Kueng, A. Neven, J. Carrasco, B. Kraus, P. Zoller, P. Calabrese, B. Vermersch, and M. Dalmonte, 
{\it Symmetry-resolved dynamical purification in synthetic quantum matter}, 
\href{https://doi.org/10.21468/SciPostPhys.12.3.106}{SciPost Phys. {\bf 12}, 106 (2022)}.

\bibitem{lr-14}
N. Laflorencie and S. Rachel, 
{\it Spin-resolved entanglement spectroscopy of critical spin chains and Luttinger liquids},
\href{http://dx.doi.org/10.1088/1742-5468/2014/11/P11013}{J. Stat. Mech. (2014) P11013}.

\bibitem{Goldstein} 
M. Goldstein and E. Sela, 
\textit{Symmetry-Resolved Entanglement in Many-Body Systems}, 
\href{https://doi.org/10.1103/PhysRevLett.120.200602}{Phys. Rev. Lett. {\bf 120}, 200602 (2018)}.

\bibitem{xavier}  
J. C. Xavier, F. C. Alcaraz, and G. Sierra, 
\textit{Equipartition of the entanglement entropy}, 
\href{https://doi.org/10.1103/PhysRevB.98.041106}{Phys. Rev. B {\bf 98}, 041106 (2018)}.

\bibitem{fis}
A. Lukin, M. Rispoli, R. Schittko, M. E. Tai, A. M. Kaufman, S. Choi, V. Khemani, J. Leonard, and M. Greiner, 
{\it Probing entanglement in a many-body localized system}, 
\href{https://dx.doi.org/10.1126/science.aau0818}{Science {\bf 364}, 6437 (2019)}.



\bibitem{vecd1-20}
A. Neven, J. Carrasco, V. Vitale, C. Kokail, A. Elben, M. Dalmonte, P. Calabrese, P. Zoller, B. Vermersch, R. Kueng, and B. Kraus, 
{\it Symmetry-resolved entanglement detection using partial transpose moments}, 
\href{https://doi.org/10.1038/s41534-021-00487-y}{Npj Quantum Inf. {\bf 7}, 152 (2021)}.

\bibitem{ahyrst-21}
D. Azses, R. Haenel, Y. Naveh, R. Raussendorf, E. Sela, and E. G. Dalla Torre, 
{\it Identification of Symmetry-Protected Topological States on Noisy Quantum Computers}, 
\href{https://doi.org/10.1103/PhysRevLett.125.120502}{Phys. Rev. Lett. {\bf 125}, 120502 (2020)}.

\bibitem{goldstein1}
E. Cornfeld, M. Goldstein, and E. Sela, 
{\it Imbalance Entanglement: Symmetry Decomposition of Negativity}, 
\href{http://dx.doi.org/10.1103/PhysRevA.98.032302} {Phys. Rev. A {\bf 98}, 032302 (2018)}.
 
\bibitem{mbc-21}
S. Murciano, R. Bonsignori, and P. Calabrese, 
{\it Symmetry decomposition of negativity of massless free fermions}, 
\href{https://doi.org/10.21468/SciPostPhys.10.5.111}{SciPost Phys. {\bf 10}, 111 (2021)}.

\bibitem{cn-21}
H.-H. Chen, 
{\it Charged R\'enyi negativity of massless free bosons},
\href{https://doi.org/10.1007/JHEP02(2022)117}{JHEP {\bf 02} (2022) 117}.

\bibitem{c-21}
H.-H. Chen, 
{\it Symmetry decomposition of relative entropies in conformal field theory},
\href{https://doi.org/10.1007/JHEP07(2021)084}{JHEP {\bf 07} (2021) 084}.

\bibitem{cc-21}
L. Capizzi and P. Calabrese, 
{\it Symmetry resolved relative entropies and distances in conformal field theory},
\href{https://doi.org/10.1007/JHEP10(2021)195}{JHEP {\bf 10} (2021) 195}.

\bibitem{uv-21}
L. Hung and G. Wong, 
{\it Entanglement branes and factorization in conformal field theory}, 
\href{https://doi.org/10.1103/PhysRevD.104.026012}{Phys. Rev. D {\bf 104}, 026012 (2021)}.

\bibitem{nonabelian-21}
P. Calabrese, J. Dubail, and S. Murciano, 
{\it Symmetry-resolved entanglement entropy in Wess-Zumino-Witten models}, 
\href{https://doi.org/10.1007/JHEP10(2021)067}{JHEP {\bf 10} (2021) 067}.

\bibitem{crc-20}
L. Capizzi, P. Ruggiero, and P. Calabrese, 
{\it Symmetry resolved entanglement entropy of excited states in a CFT}, 
\href{https://doi.org/10.1088/1742-5468/ab96b6}{J. Stat. Mech. (2020) 073101}.

\bibitem{bc-20}
R. Bonsignori and P. Calabrese, 
{\it Boundary effects on symmetry resolved entanglement}, 
\href{https://doi.org/10.1088/1751-8121/abcc3a}{J. Phys. A {\bf  54}, 015005 (2021)}.

\bibitem{eimd-20}
B. Estienne, Y. Ikhlef, and A. Morin-Duchesne
{\it Finite-size corrections in critical symmetry-resolved entanglement}, 
\href{https://doi.org/10.21468/SciPostPhys.10.3.054}{SciPost Phys. {\bf 10}, 054 (2021)}.


\bibitem{mdgc-20}
S. Murciano, G. Di Giulio, and P. Calabrese, 
{\it Entanglement and symmetry resolution in two dimensional free quantum field theories}, 
\href{https://doi.org/10.1007/JHEP08(2020)073}{JHEP {\bf 08} (2020) 073}.

\bibitem{hcc-21}
D. X. Horvath, L. Capizzi, and P. Calabrese, 
{\it U(1) symmetry resolved entanglement in free 1+1 dimensional field theories via form factor bootstrap},
\href{https://doi.org/10.1007/JHEP05(2021)197}{JHEP {\bf 05} (2021) 197}.

\bibitem{hcc-21b}
D. X. Horvath, P. Calabrese, and O. A. Castro-Alvaredo, 
{\it Branch Point Twist Field Form Factors in the sine-Gordon Model II: Composite Twist Fields and Symmetry Resolved Entanglement},
\href{https://scipost.org/10.21468/SciPostPhys.12.3.088}{SciPost Phys. {\bf 12}, 088 (2022)}.

\bibitem{dhc-20}
D. X. Horvath and P. Calabrese, {\it Symmetry resolved entanglement in integrable field theories via form factor bootstrap},
\href{https://doi.org/10.1007/JHEP11(2020)131}{JHEP {\bf 11} (2020) 131}.


\bibitem{chcc-21b}
L. Capizzi, D. X. Horvath, P. Calabrese, and O. A. Castro-Alvaredo, 
{\it Entanglement of the 3-State Potts Model via Form Factor Bootstrap: Total and Symmetry Resolved Entropies},
\href{https://doi.org/10.1007/JHEP05(2022)113}{JHEP {\bf 05} (2022) 113}.

\bibitem{bm-15}
A. Belin, L. Y. Hung, A. Maloney, S. Matsuura, R. C. Myers, and T. Sierens, 
{\it Holographic Charged Renyi Entropies}, 
\href{https://doi.org/10.1007/JHEP12(2013)059}{JHEP {\bf 12} (2013) 059}. 

\bibitem{cnn-16}
P. Caputa, M. Nozaki, and T. Numasawa, \textit{Charged entanglement entropy of local operators}, 
\href{https://doi.org/10.1103/PhysRevD.93.105032}{Phys. Rev. D \textbf{93}, 105032 (2016)}.
 
\bibitem{znm-20}
S. Zhao, C. Northe, and R. Meyer,
{\it Symmetry-Resolved Entanglement in AdS$_3$/CFT$_2$ coupled to $U(1)$ Chern-Simons Theory}, 
\href{http://dx.doi.org/10.1007/JHEP07(2021)030}{JHEP {\bf 07} (2021) 030}. 

\bibitem{wznm-20}
K. Weisenberger, S. Zhao, C. Northe, and R. Meyer,
{\it Symmetry-resolved entanglement for excited states and two entangling intervals in AdS$_3$/CFT$_2$}, 
\href{	http://dx.doi.org/10.1007/JHEP12(2021)104}{JHEP {\bf 12} (2021) 104}. 

\bibitem{znwm-22}
S. Zhao, C. Northe, K. Weisenberger, and R. Meyer,
{\it Charged Moments in $W_3$ Higher Spin Holography},
\href{https://doi.org/10.1007/JHEP05(2022)166}{JHEP {\bf 05} (2022) 166}.

\bibitem{riccarda}
R.~Bonsignori, P. Ruggiero, and P. Calabrese, 
{\it Symmetry resolved entanglement in free fermionic systems}, 
\href{https://doi.org/10.1088/1751-8121/ab4b77}{J. Phys. A {\bf 52}, 475302 (2019)}.

\bibitem{SREE2dG}
S. Fraenkel and M. Goldstein,
{\it Symmetry resolved entanglement: Exact results in 1d and beyond}, 
\href{http://dx.doi.org/10.1088/1742-5468/ab7753}{J. Stat. Mech. (2020) 033106}.

\bibitem{goldstein2}
N. Feldman and M. Goldstein, 
{\it Dynamics of Charge-Resolved Entanglement after a Local Quench}, 
\href{http://dx.doi.org/10.1103/PhysRevB.100.235146}{Phys. Rev. B {\bf 100}, 235146 (2019)}.

\bibitem{MDC-19-CTM}
S. Murciano, G. Di Giulio, and P. Calabrese, 
{\it Symmetry resolved entanglement in gapped integrable systems: a corner transfer matrix approach}, 
\href{https://dx.doi.org/10.21468/SciPostPhys.8.3.046}{SciPost Phys. {\bf 8}, 046 (2020)}.

\bibitem{ccgm-20}
P. Calabrese, M. Collura, G. Di Giulio, and S. Murciano, 
{\it Full counting statistics in the gapped XXZ spin chain},
\href{https://iopscience.iop.org/article/10.1209/0295-5075/129/60007}{EPL {\bf 129},  60007 (2020)}.

\bibitem{wv-03}
H. M. Wiseman and J. A. Vaccaro, 
{\it Entanglement of Indistinguishable Particles Shared between Two Parties},
\href{https://dx.doi.org/10.1103/PhysRevLett.91.097902}{Phys. Rev. Lett. {\bf 91}, 097902 (2003)}.

\bibitem{bhd-18}
H. Barghathi, C. M. Herdman, and A. Del Maestro, 
{\it R\'enyi Generalization of the Accessible Entanglement Entropy}, 
\href{https://doi.org/10.1103/PhysRevLett.121.150501}{Phys. Rev. Lett. {\bf 121}, 150501 (2018)}.

\bibitem{bhd-19}
H. Barghathi, E. Casiano-Diaz, and A. Del Maestro, 
{\it Operationally accessible entanglement of one dimensional spinless fermions},
\href{https://doi.org/10.1103/PhysRevA.100.022324}{Phys. Rev. A {\bf 100}, 022324 (2019)}.

\bibitem{bydm-20}
H. Barghathi, J. Yu, and A. Del Maestro
{\it Theory of noninteracting fermions and bosons in the canonical ensemble,} 
\href{https://doi.org/10.1103/PhysRevResearch.2.043206}{Phys. Rev. Res. {\bf 2}, 043206 (2020)}.

\bibitem{mrc-20}
S. Murciano, P. Ruggiero, and P. Calabrese,
 {\it Symmetry resolved entanglement in two-dimensional systems via dimensional reduction}, 
\href{https://dx.doi.org/10.1088/1742-5468/aba1e5}{J. Stat. Mech. (2020) 083102}.


\bibitem{tr-19}
M. T. Tan and S. Ryu,
{\it Particle Number Fluctuations, R\'enyi and Symmetry-resolved Entanglement Entropy 
in Two-dimensional Fermi Gas from Multi-dimensional bosonisation},
\href{https://dx.doi.org/10.1103/PhysRevB.101.235169}{Phys. Rev. B {\bf 101}, 235169 (2020)}.

\bibitem{sela-21}
Z. Ma, C. Han, Y. Meir, and E. Sela, 
{\it Symmetric inseparability and number entanglement in charge conserving mixed states}, 
\href{https://doi.org/10.1103/PhysRevA.105.042416}{Phys. Rev. A {\bf 105}, 042416 (2022)}.

\bibitem{pbc-20}
G. Parez, R. Bonsignori, and P. Calabrese, 
{\it Quasiparticle dynamics of symmetry resolved entanglement after a quench: the examples of conformal field theories and free fermions},
\href{https://doi.org/10.1103/PhysRevB.103.L041104}{Phys. Rev. B {\bf 103}, L041104  (2021)}.

\bibitem{pbc-21}
G. Parez, R. Bonsignori, and P. Calabrese, 
{\it Exact quench dynamics of symmetry resolved entanglement in a free fermion chain},
\href{https://doi.org/10.1088/1742-5468/ac21d7}{J. Stat. Mech. (2021) 093102}.

\bibitem{fg-21}
S. Fraenkel and M. Goldstein, 
{\it Entanglement Measures in a Nonequilibrium Steady State: Exact Results in One Dimension}, 
\href{https://doi.org/10.21468/SciPostPhys.11.4.085}{SciPost Phys. {\bf 11}, 085 (2021)}.

\bibitem{pbc-22}
G. Parez, R. Bonsignori, and P. Calabrese,
{\it Dynamics of charge-imbalance-resolved entanglement negativity after a quench in a free-fermion model},
\href{https://doi.org/10.1088/1742-5468/ac666c}{J. Stat. Mech. (2022) 053103}.

\bibitem{trac-20}
X. Turkeshi, P. Ruggiero, V. Alba, and P. Calabrese,
{\it Entanglement equipartition in critical random spin chains}, 
\href{https://doi.org/10.1103/PhysRevB.102.014455}{Phys. Rev. B {\bf 102}, 014455 (2020)}.

\bibitem{kusf-20}
M. Kiefer-Emmanouilidis, R. Unanyan, J. Sirker, and M. Fleischhauer, 
{\it Bounds on the entanglement entropy by the number entropy in non-interacting fermionic systems},
\href{https://doi.org/10.21468/SciPostPhys.8.6.083}{SciPost Phys. {\bf 8}, 083 (2020)}.

\bibitem{kusf-20b}
M. Kiefer-Emmanouilidis, R. Unanyan, J. Sirker, and M. Fleischhauer, 
{\it Evidence for unbounded growth of the number entropy in many-body localized phases},
\href{https://doi.org/10.1103/PhysRevLett.124.243601}{Phys. Rev. Lett. {\bf 124}, 243601 (2020)}.

\bibitem{kufs-20c}
M. Kiefer-Emmanouilidis, R. Unanyan, M. Fleischhauer, and J. Sirker,
{\it Absence of true localization in many-body localized phases}, 
\href{http://dx.doi.org/10.1103/PhysRevB.103.024203}{Phys. Rev. B {\bf 103}, 024203 (2021)}.

\bibitem{clss-19}
E. Cornfeld, L. A. Landau, K. Shtengel, and E. Sela, 
{\it Entanglement spectroscopy of non-Abelian anyons: Reading off quantum dimensions of individual anyons},
\href{http://dx.doi.org/10.1103/PhysRevB.99.115429}{Phys. Rev. B {\bf 99}, 115429 (2019)}.

\bibitem{ms-20}
K. Monkman and J. Sirker, 
{\it Operational Entanglement of Symmetry-Protected Topological Edge States},
\href{https://doi.org/10.1103/PhysRevResearch.2.043191}{Phys. Rev. Res. {\bf 2}, 043191 (2020)}.

\bibitem{as-20}
D. Azses and E. Sela, 
{\it Symmetry resolved entanglement in symmetry protected topological phases},
\href{http://dx.doi.org/10.1103/PhysRevB.102.235157}{Phys. Rev. B {\bf 102}, 235157 (2020)}.


 	
\bibitem{polym1}
A. J. Heeger, S. Kivelson, J. R. Schrieffer, and W. P. Su,
{\it Solitons in conducting polymers}, 
\href{https://doi.org/10.1103/RevModPhys.60.781}{Rev. Mod. Phys. {\bf 60}, 781 (1988)}.

\bibitem{polym2}
W. P. Su, J. R. Schrieffer, and A. J. Heeger,
{\it Solitons in Polyacetylene}, 
\href{https://doi.org/10.1103/PhysRevLett.42.1698}{Phys. Rev. Lett. {\bf 42} 1698 (1979)}.

\bibitem{ssh1}
S. Ryu and Y. Hatsugai,
{\it Topological Origin of Zero-Energy Edge States in Particle-Hole Symmetric Systems}, 
\href{https://doi.org/10.1103/PhysRevLett.89.077002}{Phys. Rev. Lett. {\bf 89}, 077002 (2002)}.
 
\bibitem{ssh2}
X.-C. Wen,
{\it Symmetry protected topological phases in non-interacting fermion systems}, 
\href{https://doi.org/10.1103/PhysRevB.85.085103}{Phys. Rev. B {\bf 85}, 085103 (2012)}.

\bibitem{Asboth} 
J. K. Asb\'oth, L. Oroszl\'any, and A. P\'alyi, 
\textit{A Short Course on Topological Insulators: Band-structure topology and edge states in one and two dimensions}, 
\href{https://doi.org/10.1007/978-3-319-25607-8}{Lect. Notes Phys. 919 (2016)}.

\bibitem{ssh3}
S. Ryu and Y. Hatsugai,
{\it Entanglement entropy and the Berry phase in the solid state}, 
\href{https://doi.org/10.1103/PhysRevB.73.245115}{Phys. Rev. B {\bf 73} 245115 (2006)}.
 
\bibitem{ssh4}
J. Sirker, M. Maiti, N.P. Konstantinidis, and N. Sedlmayr,
{\it Boundary Fidelity and Entanglement in the symmetry protected topological phase of the SSH model}
\href{http://dx.doi.org/10.1088/1742-5468/2014/10/P10032}{J. Stat. Mech. (2014) P10032}.

\bibitem{ssh5}
V. Eisler, G. Di Giulio, E. Tonni, and I. Peschel,
{\it Entanglement Hamiltonians for non-critical quantum chains,} 
\href{http://dx.doi.org/10.1088/1742-5468/abb4da}{J. Stat. Mech. 103102 (2020)}.

\bibitem{Micallo} 
T. Micallo, V. Vitale, M. Dalmonte, and P. Fromholz,
{\it Topological entanglement properties of disconnected partitions in the Su-Schrieffer-Heeger model},
\href{https://scipost.org/10.21468/SciPostPhysCore.3.2.012}{SciPost Phys. Core {\bf 3}, 012 (2020)}.

\bibitem{Zhang-17} 
S.-L. Zhang, Q. Zhou,
\textit{Two-leg Su-Schrieffer-Heeger chain with glide reflection symmetry},
\href{https://doi.org/10.1103/PhysRevA.95.061601}{Phys. Rev. A {\bf 95}, 061601(R) (2017)}.

\bibitem{PerezGonzalez1} 
B. P\'erez-Gonz\'alez, M. Bello, A. G\'omez-Le\'on, and  G. Platero,
\textit{SSH model with long-range hoppings: topology, driving and disorder},
\href{https://arxiv.org/abs/1802.03973}{arXiv:1802.03973 [cond-mat.mes-hall]}.

\bibitem{PerezGonzalez2} 
B. Pérez-González, M. Bello, A. Gómez-León, and G. Platero, 
\textit{Interplay between long-range hopping and disorder in topological systems}, 
\href{https://doi.org/10.1103/PhysRevB.99.035146}{Phys. Rev. B {\bf 99}, 035146 (2019)}.

\bibitem{Ahmadi}
N. Ahmadi, J. Abouie, and D. Baeriswyl, 
\textit{Topological and non-topological features of generalized Su-Schrieffer-Heeger models}, 
\href{https://doi.org/10.1103/PhysRevB.101.195117}{Phys. Rev. B {\bf 101}, 195117 (2020)}.

\bibitem{Hsu} 
H.-C. Hsu and T.-W. Chen, 
\textit{Topological Anderson insulating phases in the long-range Su-Schrieffer-Heeger model}, 
\href{https://doi.org/10.1103/PhysRevB.102.205425}{Phys. Rev. B {\bf 102}, 205425 (2020)}.

\bibitem{Basor} 
E. L. Basor and C. A. Tracy, 
\textit{The Fisher–Hartwig conjecture and generalizations}, 
\href{https://doi.org/10.1016/0378-4371(91)90149-7}{Physica A {\bf 177} 167 (1991)}.

\bibitem{Basor2}
E. L. Basor and K. E. Morrison,
\textit{The Fisher–Hartwig conjecture and Toeplitz eigenvalues},
\href{https://doi.org/10.1016/0024-3795(94)90187-2}{Linear Algebr. Appl. {\bf 202}, 129 (1994)}.

\bibitem{JinKorepin} 
B.-Q. Jin and V. Korepin, 
\textit{Quantum Spin Chain, Toeplitz Determinants and Fisher-Hartwig Conjecture},
\href{https://doi.org/10.1023/B:JOSS.0000037230.37166.42}{J. Stat. Phys. {\bf 116}, 79 (2004)}.

\bibitem{CalabreseEssler} 
P. Calabrese and F. H. L. Essler, 
\textit{Universal corrections to scaling for block entanglement in spin-1/2 XX chains}, 
\href{https://doi.org/10.1088/1742-5468/2010/08/P08029}{J. Stat. Mech. (2010) P08029}.

\bibitem{NIST} F. W. J. Olver, A. B. Olde Daalhuis, D. W. Lozier, B. I. Schneider, R. F. Boisvert, 
C. W. Clark, B. R. Miller, and B. V. Saunders (ed), 
\textit{NIST Digital Library of Mathematical Functions}, (Gaithersburg: National Institute of Standards and Technology) 
\href{http://dlmf.nist.gov/}{(http://dlmf.nist.gov/)}

\bibitem{Peschel} 
I. Peschel, 
\textit{Calculation of reduced density matrices from correlation functions}, 
\href{https://doi.org/10.1088/0305-4470/36/14/101}{J. Phys. A: Math. Gen. {\bf 36}, L205 (2003)}.

\bibitem{FH} 
M. E. Fisher and  R. E. Hartwig, 
\textit{Toeplitz determinants, some applications, theorems and conjectures}, 
\href{https://doi.org/10.1002/9780470143605.ch18}{Adv. Chem. Phys. {\bf 15}, 333 (1968)}.

\bibitem{Basor3}
E. L. Basor, 
\textit{A localization theorem for Toeplitz determinants}, 
\href{https://www.jstor.org/stable/24892549 }{Indiana Math. J. {\bf 28}, 975 (1979)}.

\bibitem{Widom} 
H. Widom,
\textit{Asymptotic behavior of block Toeplitz matrices and determinants},
\href{https://doi.org/10.1016/0001-8708(74)90072-3}{Adv. in Math. {\bf 13}, 284 (1974)}.

\bibitem{ccp-10} 
P. Calabrese, J. Cardy, and I. Peschel,
{\it Corrections to scaling for block entanglement in massive spin-chains},
 \href{http://dx.doi.org/10.1088/1742-5468/2010/09/P09003}{J. Stat. Mech. (2010) P09003}.

\bibitem{Ares3}
F. Ares, J. G. Esteve, and F. Falceto,
{\it Entanglement of several blocks in fermionic chains},
\href{https://doi.org/10.1103/PhysRevA.90.062321}{Phys. Rev. A {\bf 90}, 062321 (2014)}.

\bibitem{cct-12}
P. Calabrese, J. Cardy, and E. Tonni, {\it Entanglement Negativity in Quantum Field Theory}, 
\href{http://dx.doi.org/10.1103/PhysRevLett.109.130502}{Phys. Rev. Lett. {\bf 109}, 130502 (2012)}.

\bibitem{cct-13}
P. Calabrese, J. Cardy, and E. Tonni, {\it Entanglement negativity in extended systems: a field theoretical approach},
\href{http://dx.doi.org/10.1088/1742-5468/2013/02/P02008}{J. Stat. Mech. (2013) P02008}.

\bibitem{ssr-17}
H. Shapourian, K. Shiozaki, and S. Ryu, {\it Partial time-reversal transformation and entanglement negativity in fermionic systems}, 
\href{https://doi.org/10.1103/PhysRevB.95.165101}{Phys. Rev. B {\bf 95}, 165101 (2017)}.

\bibitem{sr-19}
H. Shapourian and S. Ryu, {\it  Entanglement negativity of fermions: monotonicity, separability criterion, and classification of few-mode states},
\href{https://doi.org/10.1103/PhysRevA.99.022310}{Phys. Rev. A {\bf 99}, 022310 (2019)}.

\bibitem{ryu}
H. Shapourian and S. Ryu, {\it Finite-temperature entanglement negativity of free fermions}, 
\href{https://doi.org/10.1088/1742-5468/ab11e0}{J. Stat. Mech. (2019) 043106}.

\bibitem{mvdc-22}
S. Murciano, V. Vitale, M. Dalmonte, and P. Calabrese, {\it The Negativity Hamiltonian: An operator characterization of mixed-state entanglement},
\href{https://doi.org/10.1103/PhysRevLett.128.140502}{Phys. Rev. Lett. {\bf 128}, 140502 (2022)}.

\bibitem{pcp-19}
S. Pappalardi, P. Calabrese, and G. Parisi, {\it Entanglement entropy of the long-range Dyson hierarchical model},
\href{https://doi.org/10.1088/1742-5468/ab2903}{J. Stat. Mech. (2019) 073102}.

\bibitem{m-20}
C. Monthus, {\it Properties of the simplest inhomogeneous and homogeneous Tree-Tensor-States for Long-Ranged Quantum Spin Chains with or without disorder},
\href{https://doi.org/10.1016/j.physa.2021.126040}{Physica A {\bf 576}, 126040 (2021)}.

\bibitem{Jones}
N. G. Jones,
{\it Symmetry-resolved entanglement entropy in critical free-fermion chains},
\href{https://doi.org/10.48550/arXiv.2202.11728}{arXiv:2202.11728}.

\end{thebibliography}
\end{document}